# An efficient and robust estimation of spatio-temporally distributed parameters in dynamic models by an ensemble Kalman filter

Yohei Sawada[1,2], Le Duc[1,2]

[1] Institute of Engineering Innovation, Graduate School of Engineering, the University of Tokyo, Tokyo, Japan

[2] Meteorological Research Institute, Japan Meteorological Agency, Tsukuba, Japan

Corresponding author: Y. Sawada, Institute of Engineering Innovation, the University of Tokyo, Tokyo, Japan, 2-11-6, Yayoi, Bunkyo-ku, Tokyo, Japan, yohei.sawada@sogo.t.u-tokyo.ac.jp

**Abstract**

The accuracy of Earth system models is compromised by unknown and/or unresolved dynamics, making the quantification of systematic model errors essential. While a model parameter estimation, which allows parameters to change spatio-temporally, shows promise in quantifying and mitigating systematic model errors, the estimation of the spatio-temporally distributed model parameters has been practically challenging. Here we present an efficient and practical method to estimate time-varying parameters in high-dimensional spaces. In our proposed method, Hybrid Offline and Online Parameter Estimation with ensemble Kalman filtering (HOOPE-EnKF), model parameters estimated by EnKF are constrained by results of offline batch optimization, in which the posterior distribution of model parameters is obtained by comparing simulated and observed climatological variables. HOOPE-EnKF outperforms the original EnKF in synthetic experiments using a two-scale Lorenz96 model and a simple global general circulation model. One advantage of HOOPE-EnKF over traditional EnKFs is that its performance is not greatly affected by inflation factors for model parameters, thus eliminating the need for extensive tuning of inflation factors. We thoroughly discuss the potential of HOOPE-EnKF as a practical method for improving parameterizations of process-based models and prediction in real-world applications such as numerical weather prediction.

**Plain Language Summary**

Earth system models help us understand and predict the behavior of our planet, but their accuracy is limited due to unknown or unresolved factors. Adjusting the parameters of these models based on the changing patterns in time and space can help improve their accuracy. However, this has been a challenging task. In this study, we introduce a new method called Hybrid Offline and Online Parameter Estimation with ensemble Kalman filtering (HOOPE-EnKF) that efficiently estimates these changing parameters. This method combines an existing EnKF with offline batch optimization, which compares long-term simulation with observations to fine-tune the model. Our experiments show that HOOPE-EnKF performs better than the original EnKF and is less sensitive to certain hyperparameters, reducing the need for time-consuming calibrations. HOOPE-EnKF has a great potential for improving the accuracy of models and predictions in real-world applications, like weather forecasting, and can help scientists better understand our planet's complex systems.

**Key Points**

EnKF is extended to estimate high-dimensional time-varying model parameters.

The proposed method is insensitive to the choice of covariance inflation factors.

The proposed method successfully quantifies systematic model errors from observations.

# 1. Introduction

Quantification of systematic model errors is of paramount importance in monitoring, understanding, and predicting Earth systems. The accuracy of process-based models is compromised by unknown and/or unresolved dynamics. Due to limited computational resources, many sub-grid scale parameterizations are required to solve the evolution of complex Earth systems. However, these sub-grid scale parameterizations inevitably introduce structural and parametric errors into the simulations. Although data assimilation has contributed to the prediction of Earth systems by integrating imperfect models and observations, it is essential to quantify and model the errors of process-based models to maximize the potential of data assimilation to accurately estimate the states of Earth systems.

Data-driven estimation of model errors within a data assimilation framework has been intensively investigated. For instance, Pathiraja and van Leeuwen (2022) proposed constructing a conditional model error probability density using kernel density estimation and incorporating it into an ensemble transform Kalman filter. Amemiya et al. (2023) applied a recurrent neural network to learn model errors from analysis increments estimated by local ensemble transform Kalman filter (LETKF). Tomizawa and Sawada (2021) demonstrated that emulating LETKF analysis through reservoir computing can mitigate the impact of model parameter misspecification on predictions (see also Brajard et al. (2020), Dueben and Bauer (2018), Weyn et al. (2019), and Weyn et al (2021)). All these approaches do not require knowledge of unresolved dynamics and can operate with partially observed systems. Although their real-world applications have been limited thus far, these methods are promising approaches to accurately estimate and predict the states of Earth systems by explicitly considering systematic model errors.

In the present paper, we propose estimating systematic model errors by spatio-temporally distributed model parameters. In our approach, these parameters, originally formulated as global fixed parameters in a process-based model, become spatio-temporally distributed, addressing the model bias introduced by their fixation. While this time-varying parameter estimation is particularly beneficial when the model parameters are known to be non-constant (e.g., land cover in a hydrological model; see Pathiraja et al. 2018a, 2018b), we will demonstrate and discuss the value of this approach even when the source of systematic model errors remains unclear.

In sequential data assimilation, model parameters can be estimated using an augmented state vector method, in which a parameter vector is concatenated to a state vector and then the augmented vector is updated in the analysis step (Moradkhani et al. 2005a). This simultaneous update of state variables and parameters has been successfully applied in various Earth system science domains, such as atmosphere (e.g., Kotsuki et al. 2018), land (e.g., Kurtz et al. 2016), and hydrology (e.g., Vrugt et al. 2013). However, there are few applications involving time-varying parameters in Earth system sciences, although some previous studies successfully estimated them for relatively low dimensional models (e.g., Ruiz et al. 2013a; Pathiraja et al. 2018a, 2018b; Sawada and Hanazaki 2020). The primary challenge in estimating time-varying model parameters through sequential data assimilation is maintaining the appropriate background variance of the estimated parameters. Whenever model parameters are adjusted by data assimilation, the estimated variance of the updated parameters is always smaller than that of the background. To track rapid changes in model parameters, it is necessary to maintain a relatively large spread of adjusted parameters using covariance inflation methods, which is practically difficult since ones lack knowledge of the error growth in model parameters that are originally assumed to be time-invariant. There are many works to stabilize filters to choose the appropriate inflation for parameters. Moradkhani et al. (2005b) simply added Gaussian noise, whose variance is proportional to the background variance, to the adjusted parameters of each ensemble member. Ruiz et al. (2013a) proposed an adaptive inflation method for updating ensemble parameters in LETKF. Kotsuki et al. (2018) suggested maintaining the initial predefined ensemble spread using the relaxation to prior spread method (Whitaker and Hamill 2012).

As an alternative approach to stabilize the inflation of estimated model parameters, Sawada (2022) proposed constraining model parameters estimated by sequential data assimilation to the result of offline batch optimization in which the posterior distribution of model parameters is obtained by comparing simulated and observed climatological (long-term mean) variables. The proposed method, Hybrid Offline and Online Parameter Estimation with Particle Filtering (HOOPE-PF), was successfully applied to low-dimensional systems. The performance of HOOPE-PF is insensitive to hyperparameters of an ensemble inflation method, making it a practical and efficient approach for estimating time-varying parameters. Although HOOPE-PF requires performing computationally expensive offline batch optimization, a growing number of studies have recently used machine learning-based surrogate models to accelerate this process (e.g., Dunbar et al. 2021; Zhang et al. 2020; Sawada 2020; Teixeira Parente et al. 2019; Duan et al. 2017). This makes it promising to combine online data assimilation and offline batch optimization for estimating of time-varying model parameters. However, HOOPE-PF is difficult to apply to high-dimensional Earth system models since it relies on particle filtering. In this study, we aim at applying the same concept to the widely used ensemble Kalman filter (EnKF). Although variants of EnKF have been extensively used to estimate high-dimensional unknown parameters (e.g., Kang et al. 2011; Bellsky et al. 2014; Pulido et al. 2016; Katzfuss et al. 2020; Ruckstuhl and Janjić 2020), the idea of combing offline batch optimization and EnKF to estimate spatio-temporally distributed parameters has yet to be examined. We demonstrate that our proposed EnKF-based method (i.e., HOOPE-EnKF) is useful for the estimation of systematic model errors in a two-scale Lorenz96 model and a simple atmospheric general circulation model.

## 2. Method
### 2.1. Overview of HOOPE-EnKF

The schematic of HOOPE-EnKF is shown in Figure 1. As explained in the previous section, the fundamental idea is that the analysis ensemble of model parameters in EnKF incorporates information from a predefined posterior distribution of time-invariant model parameters: $p(\widehat{\boldsymbol{\theta}}|\boldsymbol{y}(0:T)) \cong N(\boldsymbol{\theta}_c, \boldsymbol{C})$, where $\widehat{\boldsymbol{\theta}}$ is a time-invariant model parameter that can accurately simulate the climatology of the systems and $\boldsymbol{y}(0:T)$ is observation from time 0 to T. In this study, we approximate $p(\widehat{\boldsymbol{\theta}}|\boldsymbol{y}(0:T))$ as a Gaussian distribution with mean $\boldsymbol{\theta}_c$ and covariance matrix $\boldsymbol{C}$ (all parameters are assumed to be independent so that C is a diagonal matrix). In the next sub-section, we will explain how to obtain $p(\widehat{\boldsymbol{\theta}}|\boldsymbol{y}(0:T))$ by offline batch optimization. In Section 2.3, our sequential data assimilation method, LETKF, will be formulated. In the section 2.4, we will introduce HOOPE-EnKF as a simple extension of LETKF.

### 2.2. Offline batch optimization

A discrete state-space dynamic system is written as:

$$\boldsymbol{x}(t) = f\left(\boldsymbol{x}(t-1), \widehat{\boldsymbol{\theta}}, \boldsymbol{u}(t-1)\right) + \boldsymbol{q}(t-1) \qquad (1)$$

$$\boldsymbol{y}^o(t) = H\big(\boldsymbol{x}(t)\big) + \boldsymbol{r}(t) \qquad (2)$$

where $\boldsymbol{x}(t)$ are the state variables at time t, $\widehat{\boldsymbol{\theta}}$ is the time-invariant model parameters which are the originally defined parameters in process-based models (see Section 1), $\boldsymbol{u}(t)$ is the external forcing at time t, and $\boldsymbol{q}(t)$ is the noise process representing model error at time t. $f()$ denotes the dynamic model. $y^o(t)$ is the observation at time t, $H$ is the observation operator, and $r(t)$ is the noise process representing observation error. In this study, a simulated climatological index, $\gamma^f$, is calculated from the long-term timeseries of the simulated observation $y^f(0:T)$

$$\gamma^f = g(\hat{\theta}) \tag{3}$$

where $\gamma^f$ is the simulated climatological index and $g()$ is the forward map. Note that $\gamma^f$ should not be a function of the system's initial conditions since the targeted climatological index for optimizing time-invariant model parameters should not be flow-dependent. The purpose of the offline batch optimization in this study is to obtain the posterior distribution of $\hat{\theta}$, or $p(\hat{\theta}|y(0:T))$, based on the simulated climatological index $\gamma^f$ and the observed climatological index $\gamma^o$.

The fully Bayesian inference using Equation 3 is impractical because $g(.)$ is computationally expensive. To efficiently estimate $p(\hat{\theta}|y(0:T))$, a surrogate model of $g(.)$ was developed. First, ensemble members of model parameters were generated from a parameter space. Second, the long-term integration of the dynamic model with these ensemble members was performed in parallel. Third, the statistical surrogate model was constructed from the data of this ensemble simulation by Gaussian process regression (Rasmussen and Williams 2006). There are two ways to replace the full process-based model integration with the statistical surrogate model.

$$\gamma^f = g^{(s)}(\hat{\theta}) \tag{4a}$$

$$\|\gamma^f - \gamma^o\|^2 = g^{(s)}(\hat{\theta}) \tag{4b}$$

where $g^{(s)}(.)$ is the computationally cheap surrogate model of $g(.)$. We sampled $\hat{\theta}$ and performed ensemble simulation with the sampled parameters. Then, we fit the sampled relationship between $\hat{\theta}$ and our target metrices using Gaussian process regression with the Matern kernel. While the long-term integration of the dynamic model with input data such as initial and boundary conditions and external forcings is needed to obtain the model's climatology and evaluate $g(\hat{\theta})$, the trained surrogate model directly estimates the relationship between model behaviors and parameters without performing the long-term integration. Equation (4a) approximates the simulated climatology by Gaussian process regression while we directly approximate the square difference between simulated and observed climatological index in Equation (4b). Equation (4b) is more suitable if the dimension of $\gamma^f$ is high. When we fit the square difference between simulation and observation, our dependent variable in the regression is a scaler, which makes the construction of a surrogate model simpler. We adopted (4a) in the two-scale

Lorenz 96 model experiment, while (4b) is adopted in the experiment with a simple atmospheric model.

We applied the Metropolis-Hastings algorithm (Hastings, 1970) to draw samples from $p(\widehat{\boldsymbol{\theta}}|\boldsymbol{y}(0:T))$. We assumed a prior distribution as a trimmed uniform distribution with predefined maximum and minimum values of parameters. Please refer to Sawada (2022) for details of this sampler (see also section 3 for the implementation of the sampler). Although Sawada (2022) directly used $p(\widehat{\boldsymbol{\theta}}|\boldsymbol{y}(0:T))$, HOOPE-EnKF needs to approximate it to a Gaussian distribution. In this study, we simply calculate mean and variance of samples from the Metropolis-Hastings algorithm to get $N(\boldsymbol{\theta}_c, \boldsymbol{C})$. A transformation of variables needs to be performed if $p(\widehat{\boldsymbol{\theta}}|\boldsymbol{y}(0:T))$ deviates from the Gaussian distribution. If $p(\widehat{\boldsymbol{\theta}}|\boldsymbol{y}(0:T))$ substantially deviates from the Gaussian distribution and the transformation does not work, it is intrinsically difficult to apply HOOPE-EnKF, which is the limitation of our method.

It should be noted that there are many other practically promising approaches to obtain $N(\boldsymbol{\theta}_c, \boldsymbol{C})$. For instance, iterative ensemble smoothers (Evensen, 2018) and their extensions (e.g., Cleary et al. 2021) are effective in obtaining $N(\boldsymbol{\theta}_c, \boldsymbol{C})$ with a relatively smaller number of iterations. If a dynamic model has already been extensively calibrated using a targeted climatological index, the expert knowledge of modelers can be beneficial to directly specify $N(\boldsymbol{\theta}_c, \boldsymbol{C})$ (e.g., using their default settings as $\boldsymbol{\theta}_c$).

## 2.3. Local Ensemble Transform Kalman Filter (LETKF)

In LETKF and any other ensemble Kalman filters, the analysis probability distribution is entirely determined by its first and second moments which can be obtained from the minimum point and its associated Hessian matrix of the following cost function J:

$$J = \frac{1}{2}(\boldsymbol{x} - \overline{\boldsymbol{x}}^b)^T(\boldsymbol{B}_x)^{-1}(\boldsymbol{x} - \overline{\boldsymbol{x}}^b) + \frac{1}{2}(\boldsymbol{y}^o - H(\boldsymbol{x}))^T \boldsymbol{R}^{-1}(\boldsymbol{y}^o - H(\boldsymbol{x})) \qquad (5)$$

where $\overline{\boldsymbol{x}}^b$ is the background ensemble mean of state variables, $\boldsymbol{B}_x$ is the background error covariance of $\boldsymbol{x}$. $\boldsymbol{R}$ is the observation error covariance. The first and second terms in Equation 5 are called background and observation terms, respectively.

To estimate parameters $\boldsymbol{\theta}$, state augmentation is effective as discussed in Section 1. For the augmented vector, the cost function can be rewritten as:

$$J = \frac{1}{2}\begin{pmatrix}x - \bar{x}^b \\ \theta - \bar{\theta}^b\end{pmatrix}^T \begin{pmatrix}B_x & B_{x\theta} \\ B_{\theta x} & B_\theta\end{pmatrix}^{-1} \begin{pmatrix}x - \bar{x}^b \\ \theta - \bar{\theta}^b\end{pmatrix} + \frac{1}{2}(y^o - h(x))^T R^{-1}(y^o - h(x)) \quad (6)$$

where $\bar{\theta}^b$ is the background ensemble mean of model parameters, $B_\theta$ is the background error covariance of $\theta$, and $B_{x\theta}$ and $B_{\theta x}$ are the background error covariances showing the correlation between state variables and model parameters, which is the key for transferring information from observable variables to model parameters. Note that these model parameters, $\theta$, in Equation 6 are no longer time-invariant ones, so that they are updated in the analysis step.

When we redefine the augmented state vector $x_{aug} = \begin{pmatrix}x \\ \theta\end{pmatrix}$, the LETKF's analysis update is as follows (see Hunt et al. 2007 for the complete description):

$$w^a = \widetilde{P^a}(Y^b)^T R^{-1}(y^o - \bar{y}^b) \quad (7)$$
$$\widetilde{P^a} = [(k-1)I + (Y^b)^T R^{-1} Y^b]^{-1} \quad (8)$$
$$\bar{x}^a_{aug} = \bar{x}^b_{aug} + X^b w^a \quad (9)$$

where $y^o$ and $\bar{y}^b$ are the observation and the ensemble mean of the simulated observation, respectively. k is the ensemble size. The *i*th columns of $Y^b$ and $X^b$ are $y^{b(i)} - \bar{y}^b$ and $x^{b(i)}_{aug} - \bar{x}^b_{aug}$, respectively. $y^{b(i)}$ is the simulated observation of the *i*th ensemble member, and $x^{b(i)}_{aug}$ is the augmented state vector $x_{aug}$ of the *i*th background ensemble member. $\bar{x}^a_{aug}$ and $\bar{x}^b_{aug}$ are the background and analysis ensemble means of the augmented state vector $x_{aug}$, respectively. The perturbations of the analysis ensemble members can be found by:

$$W^a = [(k-1)\widetilde{P^a}]^{\frac{1}{2}} \quad (10)$$
$$X^a = X^b W^a \quad (11)$$

The *i*th column of $X^a$ is $x^{a(i)}_{aug} - \bar{x}^a_{aug}$, where $x^{a(i)}_{aug}$ is the augmented state vector of the *i*th analysis ensemble member. From $X^a$, we can immediately obtain the analysis ensemble members of both model state variables and parameters.

As we will show in the following sections, HOOPE-EnKF modifies Equation 6 and does not necessarily change the method for minimizing the cost function. Our proposed method and the results presented in this paper are applicable to other flavors of EnKF.

## 2.4. Hybrid Offline and Online Parameter Estimation with Ensemble Kalman Filtering (HOOPE-EnKF)

As discussed in Section 2.1, the analysis ensemble of model parameters in HOOPE-EnKF incorporates information from a posterior distribution of time-invariant model parameters: $p(\widehat{\boldsymbol{\theta}}|\boldsymbol{y}(0:T)) \cong N(\boldsymbol{\theta}_c, \boldsymbol{C})$. Since the background term in Equation 6 can be recognized as a regularization term, it is straightforward to incorporate $N(\boldsymbol{\theta}_c, \boldsymbol{C})$ as an additional regularization term:

$$J = \frac{1}{2}\begin{pmatrix}\boldsymbol{x} - \overline{\boldsymbol{x}}^b \\ \boldsymbol{\theta} - \overline{\boldsymbol{\theta}}^b\end{pmatrix}^T \begin{pmatrix}\boldsymbol{B}_x & \boldsymbol{B}_{x\theta} \\ \boldsymbol{B}_{\theta x} & \boldsymbol{B}_\theta\end{pmatrix}^{-1} \begin{pmatrix}\boldsymbol{x} - \overline{\boldsymbol{x}}^b \\ \boldsymbol{\theta} - \overline{\boldsymbol{\theta}}^b\end{pmatrix} + \frac{1}{2}(\boldsymbol{y}^o - h(\boldsymbol{x}))^T \boldsymbol{R}^{-1}(\boldsymbol{y}^o - h(\boldsymbol{x})) + \frac{1}{2}(\boldsymbol{\theta} - \boldsymbol{\theta}_c)^T \boldsymbol{C}^{-1}(\boldsymbol{\theta} - \boldsymbol{\theta}_c) \quad (12)$$

Since $N(\boldsymbol{\theta}_c, \boldsymbol{C})$ represent the "climatology" of the parameters, we call the third term of the right-hand side of Equation 12 the climatological term. To solve Equation 12 with the standard EnKF, we will convert this new form to the standard form of the cost function (Equation 6). We identified two approaches to accomplish this task.

### 2.4.1. Pseudo observation (PSO) approach

As a first approach, we combined the observation and climatological terms in Equation 12 to get a more compact form:

$$J = \frac{1}{2}\begin{pmatrix}\boldsymbol{x} - \overline{\boldsymbol{x}}^b \\ \boldsymbol{\theta} - \overline{\boldsymbol{\theta}}^b\end{pmatrix}^T \begin{pmatrix}\boldsymbol{B}_x & \boldsymbol{B}_{x\theta} \\ \boldsymbol{B}_{\theta x} & \boldsymbol{B}_\theta\end{pmatrix}^{-1} \begin{pmatrix}\boldsymbol{x} - \overline{\boldsymbol{x}}^b \\ \boldsymbol{\theta} - \overline{\boldsymbol{\theta}}^b\end{pmatrix} + \frac{1}{2}\begin{pmatrix}\boldsymbol{y}^o - h(\boldsymbol{x}) \\ \boldsymbol{\theta}_c - \boldsymbol{\theta}\end{pmatrix}^T \begin{pmatrix}\boldsymbol{R} & 0 \\ 0 & \boldsymbol{C}\end{pmatrix}^{-1} \begin{pmatrix}\boldsymbol{y}^o - h(\boldsymbol{x}) \\ \boldsymbol{\theta}_c - \boldsymbol{\theta}\end{pmatrix} \quad (13)$$

This form essentially has the same structure as Equations 5 and 6 by defining the "augmented observations" $\boldsymbol{y}_{aug} = \begin{pmatrix}\boldsymbol{y}^o \\ \boldsymbol{\theta}_c\end{pmatrix}$, the associated observation operator $h_{aug}(\boldsymbol{x}, \boldsymbol{\theta}) = \begin{pmatrix}h(\boldsymbol{x}) \\ \boldsymbol{\theta}\end{pmatrix}$, and the new observation error covariance $\boldsymbol{R}_{aug} = \begin{pmatrix}\boldsymbol{R} & 0 \\ 0 & \boldsymbol{C}\end{pmatrix}$. The implication of Equation 13 is that we can solve Equation 12 with the standard EnKF by assimilating $\boldsymbol{\theta}_c$ into the system with variance $\boldsymbol{C}$. We call this approach "pseudo observation" since we used $\boldsymbol{\theta}_c$ as a pseudo observation of model parameters.

It is important to discuss how to implement inflation and localization in the PSO approach. In this study, we used prior multiplicative inflation which involves multiplying the background covariances with inflation factors. We simply applied the factors to the background perturbations in the PSO approach. In LETKF, observation localization is used. In this study, we let pseudo-observations of $\boldsymbol{\theta}_c$ to impact only the associate parameter variables. See also Section 3 for details of the implementation of inflation and localization.

### 2.4.2. Regression to climatology (RTC) approach

The second idea is to combine the background and climatological terms in Equation 12 to form a general quadratic form

$$J = \frac{1}{2}\begin{pmatrix} x - \bar{x}^b \\ \theta - \bar{\theta}^b \end{pmatrix}^T \begin{pmatrix} B_x & B_{x\theta} \\ B_{\theta x} & B_\theta \end{pmatrix}^{-1} \begin{pmatrix} x - \bar{x}^b \\ \theta - \bar{\theta}^b \end{pmatrix} + \frac{1}{2}(\theta - \theta_c)^T C^{-1}(\theta - \theta_c)$$

$$= \frac{1}{2}\begin{pmatrix} x - \tilde{x}^b \\ \theta - \tilde{\theta}^b \end{pmatrix}^T \begin{pmatrix} \tilde{B}_x & \tilde{B}_{x\theta} \\ \tilde{B}_{\theta x} & \tilde{B}_\theta \end{pmatrix}^{-1} \begin{pmatrix} x - \tilde{x}^b \\ \theta - \tilde{\theta}^b \end{pmatrix} + const \quad (14)$$

Due to the presence of $B_{x\theta}$ and $B_{\theta x}$, it is relatively complicated to find the analytic forms of the tilde variables on the right-hand side of (14). The detailed derivation is given in Appendix A and the result is reproduced here

$$\begin{pmatrix} \tilde{x}^b \\ \tilde{\theta}^b \end{pmatrix} = \begin{pmatrix} \bar{x}^b + B_{x\theta}[C + B_\theta]^{-1}(\theta_c - \bar{\theta}^b) \\ C[C + B_\theta]^{-1}\bar{\theta}^b + B_\theta[C + B_\theta]^{-1}\theta_c \end{pmatrix} \quad (15a)$$

$$\begin{pmatrix} \tilde{B}_x & \tilde{B}_{x\theta} \\ \tilde{B}_{\theta x} & \tilde{B}_\theta \end{pmatrix} = \begin{pmatrix} B_x - B_{x\theta}[C + B_\theta]^{-1}B_{\theta x} & B_{x\theta}[C + B_\theta]^{-1}C \\ C[C + B_\theta]^{-1}B_{\theta x} & [C^{-1} + B_\theta^{-1}]^{-1} \end{pmatrix} \quad (15b)$$

In EnKF, we approximate the pdf associated with the updated background term (14) by ensemble samples. In principle, we can draw new samples from the updated covariance (15b) by taking an appropriate square root of this matrix. However, this strategy is very ineffective in practice due to the high computation cost in estimating such a square root. Note that the computational cost of RTC should be comparable to that of PSO, otherwise we will stick to the PSO approach due to its fast computation. Now note that the updated moments $\tilde{\theta}^b$ and $\tilde{B}_\theta$ does not contain any terms related to $x$, this suggests an approximation that treats $x$ and $\theta$ independently. Thus, our strategy is to draw samples of $\theta$ from the updated marginal distribution $\mathcal{N}(\tilde{\theta}^b, \tilde{B}_\theta)$, which is relatively easy since $\tilde{B}_\theta$ is usually a diagonal matrix, while approximating the remaining moments.

Recall that in the PSO method, since the background term is not modified, it is straightforward to use the same samples $\{\theta^{b(i)}\}$ representing $\mathcal{N}(\bar{\theta}^b, B_\theta)$ as in the standard case of Equation 6. However, in the current method, since we update $\mathcal{N}(\bar{\theta}^b, B_\theta)$ to $\mathcal{N}(\tilde{\theta}^b, \tilde{B}_\theta)$, it is necessary to update $\{\theta^{b(i)}\}$ to new samples $\{\tilde{\theta}^{b(i)}\}$ that conform with the updated pdf. We propose using optimal transport to move $\{\theta^{b(i)}\}$ drawn from $\mathcal{N}(\bar{\theta}^b, B_\theta)$ to $\{\tilde{\theta}^{b(i)}\}$ that faithfully represent $\mathcal{N}(\tilde{\theta}^b, \tilde{B}_\theta)$. Because $\mathcal{N}(\overline{\boldsymbol{\theta}}^b, \mathbf{B}_\theta)$ and $\mathcal{N}(\widetilde{\boldsymbol{\theta}}^b, \widetilde{\mathbf{B}}_\theta)$ are the Gaussian distributions, the optimal transport map has an analytical form (Peyré and Cuturi, 2019):

$$\widetilde{\boldsymbol{\theta}}^{b(i)} = T(\boldsymbol{\theta}^{b(i)}) = \widetilde{\boldsymbol{\theta}}^b + \boldsymbol{B}_\theta^{-\frac{1}{2}} \left( \boldsymbol{B}_\theta^{\frac{1}{2}} \widetilde{\boldsymbol{B}}_\theta \boldsymbol{B}_\theta^{\frac{1}{2}} \right)^{\frac{1}{2}} \boldsymbol{B}_\theta^{-\frac{1}{2}} (\boldsymbol{\theta}^{b(i)} - \overline{\boldsymbol{\theta}}^b) \tag{16}$$

where all matrix square roots denote symmetric positive-definite square roots.

The remaining task now is how we draw new samples $\{\widetilde{x}^{b(i)}\}$ that approximates moments related to $x$ in (15a) and (15b). Since, it is important to maintain dynamical balance in each member, we simply keep all $\{x^{b(i)}\}$ intact by setting $\widetilde{x}^{b(i)} = x^{b(i)}$. The transformation (16) when $C, B_\theta$ are diagonal matrices has the following form

$$\widetilde{\boldsymbol{\theta}}^{b(i)} - \widetilde{\boldsymbol{\theta}}^b = \boldsymbol{C}^{1/2}[\boldsymbol{C} + \boldsymbol{B}_\theta]^{-1/2}(\boldsymbol{\theta}^{b(i)} - \overline{\boldsymbol{\theta}}^b) \tag{17}$$

These updated perturbations enable us to check how our approximation alters the analytic form of the updated background term

$$\begin{pmatrix} \widetilde{x}_{app}^b \\ \widetilde{\boldsymbol{\theta}}^b \end{pmatrix} = \begin{pmatrix} \overline{x}^b \\ \boldsymbol{C}[\boldsymbol{C} + \boldsymbol{B}_\theta]^{-1}\overline{\boldsymbol{\theta}}^b + \boldsymbol{B}_\theta[\boldsymbol{C} + \boldsymbol{B}_\theta]^{-1}\boldsymbol{\theta}_c \end{pmatrix} \tag{18a}$$

$$\begin{pmatrix} \widetilde{\boldsymbol{B}}_{xapp} & \widetilde{\boldsymbol{B}}_{x\theta app} \\ \widetilde{\boldsymbol{B}}_{\theta x app} & \widetilde{\boldsymbol{B}}_\theta \end{pmatrix} = \begin{pmatrix} \boldsymbol{B}_x & \boldsymbol{B}_{x\theta}[\boldsymbol{C} + \boldsymbol{B}_\theta]^{-1/2}\boldsymbol{C}^{1/2} \\ \boldsymbol{C}^{1/2}[\boldsymbol{C} + \boldsymbol{B}_\theta]^{-1/2}\boldsymbol{B}_{\theta x} & [\boldsymbol{C}^{-1} + \boldsymbol{B}_\theta^{-1}]^{-1} \end{pmatrix} \tag{18b}$$

where all quantities with the suffice app denote the approximated moments. Comparing the approximated form (18) and the exact form (15), it is easy to see that our approximation retains the original uncertainty of the state $\boldsymbol{B}_x - \boldsymbol{B}_{x\theta}(\boldsymbol{C} + \boldsymbol{B}_\theta)^{-1}\boldsymbol{B}_{\theta x} \to \boldsymbol{B}_x$ while slightly increases the cross-correlations $\boldsymbol{C}(\boldsymbol{C} + \boldsymbol{B}_\theta)^{-1}\boldsymbol{B}_{\theta x} \to \boldsymbol{C}^{1/2}(\boldsymbol{C} + \boldsymbol{B}_\theta)^{-1/2}\boldsymbol{B}_{\theta x}$. This entails that a certain inflation has been introduced to analysis increments and analysis perturbations of parameters due to this increase.

This approximation leads to an approximated form of Equation 12 under the standard form of Equation 6:

$$J = \frac{1}{2} \begin{pmatrix} x - \overline{x}^b \\ \boldsymbol{\theta} - \widetilde{\boldsymbol{\theta}}^b \end{pmatrix}^T \begin{pmatrix} \boldsymbol{B}_x & \boldsymbol{B}_{x\theta}[\boldsymbol{C} + \boldsymbol{B}_\theta]^{-1/2}\boldsymbol{C}^{1/2} \\ \boldsymbol{C}^{1/2}[\boldsymbol{C} + \boldsymbol{B}_\theta]^{-1/2}\boldsymbol{B}_{\theta x} & \widetilde{\boldsymbol{B}}_\theta \end{pmatrix}^{-1} \begin{pmatrix} x - \overline{x}^b \\ \boldsymbol{\theta} - \widetilde{\boldsymbol{\theta}}^b \end{pmatrix}$$
$$+ \frac{1}{2}(y^o - h(x))^T \boldsymbol{R}^{-1}(y^o - h(x)) \tag{19}$$

It is possible to replace the first moment $\overline{x}^b$ in (19) by the exact one $\widetilde{x}^b$ in (15a) by simply adding $\boldsymbol{B}_{x\theta}[\boldsymbol{C} + \boldsymbol{B}_\theta]^{-1}(\boldsymbol{\theta}_c - \overline{\boldsymbol{\theta}}^b)$ to all $\{x_{b(i)}\}$. However, this may spoil dynamical balance in each member, therefore, we do not apply this kind of bias correction in this study. We call this approach the regression to climatology (RTC) since the prior pdf $\mathcal{N}(\overline{\boldsymbol{\theta}}^b, \mathbf{B}_\theta)$ is replaced with its regressed version $\mathcal{N}(\widetilde{\boldsymbol{\theta}}^b, \widetilde{\mathbf{B}}_\theta)$.

Compared to the PSO method, it is subtle to apply prior multiplicative inflation to Equation 19. When the multiplicative inflation factor for parameters is defined as $\rho_\theta$, the correct inflated covariance of parameter, $\widetilde{\mathbf{B}}_{\theta inf}$, should not be taken as $\rho_\theta \widetilde{\mathbf{B}}_\theta$. It should be derived from Equation 16:

$$\widetilde{\mathbf{B}}_{\theta inf}^{-1} = \mathbf{C}^{-1} + \rho_\theta^{-1} \mathbf{B}_\theta^{-1} \tag{20}$$

which entails:

$$\widetilde{\boldsymbol{\theta}}_{inf}^b = \widetilde{\mathbf{B}}_{\theta inf} \mathbf{C}^{-1} \boldsymbol{\theta}_c + \rho_\theta^{-1} \widetilde{\mathbf{B}}_{\theta inf} \mathbf{B}_\theta^{-1} \overline{\boldsymbol{\theta}}^b \tag{21}$$

Where $\widetilde{\boldsymbol{\theta}}_{binf}$ is the inflated version of the background ensemble mean of parameters after the optimal transportation. As the inflated background perturbations, we can obtain:

$$\widetilde{\boldsymbol{\theta}}_{inf}^{b(i)} - \widetilde{\boldsymbol{\theta}}_{inf}^b = \mathbf{B}_\theta^{-\frac{1}{2}} \left( \mathbf{B}_\theta^{\frac{1}{2}} \widetilde{\mathbf{B}}_{\theta inf} \mathbf{B}_\theta^{\frac{1}{2}} \right)^{\frac{1}{2}} \mathbf{B}_\theta^{-\frac{1}{2}} \left( \boldsymbol{\theta}^{b(i)} - \overline{\boldsymbol{\theta}}^b \right) \tag{22}$$

The RTC approach appears more complicated and requires higher computational costs than the PSO approach. However, the computation can be significantly reduced when all parameters are assumed to be uncorrelated. In this case, the inflated background perturbation for each parameter takes a simple form:

$$\tilde{\theta}_{inf}^{b(i)} - \tilde{\theta}_{inf}^b = \frac{\rho_\theta^{\frac{1}{2}} \sigma_c}{\sqrt{\sigma_c^2 + \rho_\theta \sigma_b^2}} \left( \theta^{b(i)} - \overline{\theta}^b \right) \tag{23}$$

where $\sigma_c^2, \sigma_b^2$ are climatological and background error variances of the parameter $\theta$, respectively. We can consider the factor $\rho_\theta \sigma_c^2 / (\sigma_c^2 + \rho_\theta \sigma_b^2)$ as the effective inflation factor in the RTC method. The ensemble spread is inflated when $\rho_\theta > \frac{1}{1 - \sigma_b^2/\sigma_c^2}$. Since we usually have $\sigma_b \ll \sigma_c$, the ensemble spread is expected to be increased by Equation (23). Note that we still need to compute $\tilde{\theta}_{inf}^b$ since it is required in estimating the posterior mean of $\theta$:

$$\tilde{\theta}_{inf}^b = \frac{\rho_\theta \sigma_b^2}{\sigma_c^2 + \rho_\theta \sigma_b^2} \theta_c + \frac{\sigma_c^2}{\sigma_c^2 + \rho_\theta \sigma_b^2} \overline{\theta}^b \tag{24}$$

When $\rho_\theta \to \infty$, $\tilde{\theta}_{inf}^b \to \theta_c$ and $\rho_\theta \sigma_c^2 / (\sigma_c^2 + \rho_\theta \sigma_b^2) \to \sigma_c^2 / \sigma_b^2$. Equations 23 and 24 thus explain the actual effect of $\rho_\theta$. We nudge background perturbations toward the climatological pdf $N(\boldsymbol{\theta}_c, \mathbf{C})$. In the limit $\rho_\theta \to \infty$, we replace all background samples by their climatological counterparts, ensuring the stable performance of our method.

## 3. Experiment design
### 3.1. Two-scale Lorenz 96 model

We tested HOOPE-EnKF using the two-scale Lorenz 96 model:

$$\frac{dX[k]}{dt} = -X[k-1](X[k-2] - X[k+1]) - X[k] + S + U[k]; k \in \{1,..,N_x\} \quad (25)$$

$$\xi \frac{dV[l,k]}{dt} = -V[l+1,k](V[l+2,k] - V[l-1,k]) - V[l,k] + h_z X[k]; l \in \{1,..,N_z\} \quad (26)$$

$$U[k] = \frac{h_x}{N_z} \sum_{l=1}^{N_z} V[l,k] \quad (27)$$

where $X[k]$ is the state variables of the system evolving on a coarse spatial scale with low-frequency and large amplitude fluctuations at the $k$th grid point. $X[k]$ is coupled to small-scale variables $V[l,k]$ which evolve with high frequency and small amplitude fluctuations. Equation 27 shows the sub-grid scale tendency, which represents the effect of the unresolved small-scale dynamics on the large-scale phenomena. The system has periodic boundary conditions, so that $X[k] = X[k+N_x], V[l,k] = V[l,N_z]$, and $V[l+N_z,k] = V[l,k+1]$. This testbed has been widely used in previous studies on parameter estimation and uncertainty quantification (e.g., Pathiraja and van Leeuwen 2022; Amemiya et al. 2023; Pulido et al. (2018)).

The dimension of the large-scale system, $N_x$, was set to 9, while that of the small-scale system, $N_z$, was set to 20. The external forcing, $S$, was set to 14. The time scale separation, $\xi$, was set to 0.7. We set $h_x$ and $h_z$ to -2 and 1, respectively. These settings are identical to those used by Pathiraja and van Leeuwen (2022).

In our synthetic experiment, the true dynamics are represented by Equations 25-27. We assumed that our available forecast model is the single-scale Lorenz96 model:

$$\frac{dX[k]}{dt} = -X[k-1](X[k-2] - X[k+1]) - X[k] + F[k]; k \in \{1,..,N_x\} \quad (28)$$

where F[k] is a model parameter. To make Equation 28 represent the true dynamics, $F[k] = S + U[k]$, which implies that F[k] should be spatio-temporally distributed parameters. The objective of our experiment is to estimate $F[k] = S + U[k]$ from sparsely distributed observations without any knowledge of S and Equations 26-27.

Both the two-scale and single-scale Lorenz96 models were numerically solved by a fourth-order Runge-Kutta method with a time step of $\Delta t = 0.0005$. Following Pathiraja and van Leeuwen (2022) and Amemiya et al. (2023), we used model time unit (MTU), where $1MTU = \frac{1}{\Delta t}$. We used the 7200 MTUs timeseries of the two-scale Lorenz96 model as the truth. Observations of the resolved large-scale variables were generated by adding the Gaussian white noise to the true variables. We assumed that $X[k]$ can be observed at $k = 1,2,5,6$ every 0.05 MTU. Note that Pathiraja and van Leeuwen (2022) also assumed that the system can be observed at the same grid points. The standard deviation of the Gaussian white noise was set to 0.1, so that we assumed that observation error is 0.1, which is much larger than that assumed by Pathiraja and van Leeuwen (2022).

In the offline batch optimization, we assumed that $F[k]$ is a global time-invariant parameter, $\hat{F}$. First, we generated 100 ensemble members of $\hat{F}$ from $\hat{F} = 0$ to $\hat{F} = 30$ with constant intervals and then ran the single-scale Lorenz96 model for 28800 MTUs. Second, we calculated a climatological index from the ensemble simulations, $\gamma^f$ in Equation 3. We used autocorrelation of $X[k]$ for $\gamma^f$. We calculated three autocorrelations with time lags of 2, 3, and 4 observation intervals, which correspond to 0.1, 0.15, and 0.2 MTU, respectively. Note that the dimension of $\gamma^f$ is 3. The autocorrelations were calculated for the last 2000 MTUs. Third, we smoothed the sampled relationship between $\hat{F}$ and $\gamma^f$ by the Gaussian process regression to construct the surrogate model shown in Equation 4. Fourth, by comparing $\gamma^f$ estimated by the surrogate model with $\gamma^o$ (i.e., autocorrelation estimated from observation), we drew the samples of $\hat{F}$ by the Metropolis-Hasting algorithm. Because of the small observation error, the effect of added noises to the truth on autocorrelation was negligible. Discarding the first 100,000 samples as the spin-up period, the remaining 400,000 samples are used to estimate $N(\boldsymbol{\theta}_c, \boldsymbol{C})$. Note that the offline batch optimization gives a single sample mean and variance although $N(\boldsymbol{\theta}_c, \boldsymbol{C})$ includes information of parameters in all grid points. These single mean and variance are applied to all elements of the vector $\boldsymbol{\theta}_c$ and the diagonal part of $\boldsymbol{C}$. For details on the implementation of the offline batch optimization, see Appendix B.

We implemented three types of LETKF. First, we applied the vanilla LETKF without using the climatology of parameters, $N(\boldsymbol{\theta}_c, \boldsymbol{C})$. In this case, we simply solved Equation 6 with the augmented state vector to estimate both state variables and spatio-temporally distributed parameters. This experiment is referred to as the NOHOOPE experiment. Our second and third experiments are HOOPE-EnKF with the PSO approach (Section 2.4.1) and the RTC approach (Section 2.4.2), respectively. They are called the HOOPE-EnKF-PSO and HOOPE-EnKF-RTC experiments. We implemented three LETKFs with ensemble sizes of 10, 20, and 40.

We performed prior multiplicative inflation for all three LETKFs. For the NOHOOPE and HOOPE-EnKF-PSO experiments, the ensemble perturbations were inflated as follows:

$$x_{inf}^{b(i)} - \bar{x}^b = \rho_x^{\frac{1}{2}}(x^{b(i)} - \bar{x}^b) \quad (29)$$

$$\theta_{inf}^{b(i)} - \bar{\theta}^b = \rho_\theta^{\frac{1}{2}}(\theta^{b(i)} - \bar{\theta}^b) \quad (30)$$

where $\rho_x, \rho_\theta \geq 1$ are multiplicative inflation factors for state variables and parameters, respectively. For the HOOPE-EnKF-RTC experiment, Equation 29 was used to inflate the ensemble perturbations of state variables, while Equations 23-24 were used for inflating parameter perturbations. Please note that we used different multiplicative inflation factors for state variables and parameters. We conducted three LETKFs (i.e. NOHOOPE, HOOPE-EnKF-PSO, and HOOPE-EnKF-RTC) with $31 \times 31 = 961$ combinations of $\rho_x$ and $\rho_\theta$, with ranges [1.05,2.55] and [1.05,7.05], respectively. In addition, we performed three LETKFs with adaptive inflation factors. We used the Gaussian approach proposed by Miyoshi (2011) to adaptively estimate the multiplicative inflation factors. In this case, different multiplicative inflation factors estimated adaptively were used for state variables and parameters in different grid points.

For the observation localization of LETKF in this study, the inverse of observation covariance matrix $R^{-1}$ was multiplied by the localization function $L(r)$:

$$L(r) = \begin{cases} exp\left(-\frac{1}{2}\left(\frac{r}{\sigma}\right)^2\right) & if\ r < 2\sqrt{10/3}\sigma \\ 0 & if\ r \geqq 2\sqrt{10/3}\sigma \end{cases} \quad (31)$$

where $r$ is the distance from an observation location to an analyzed grid point. The use of tapering functions has been widely adopted in LETKF applications since Miyoshi and Amane (2006) proposed them. $\sigma$ was set to 3.0. The choice of $\sigma$ does not qualitatively change our conclusions (not shown). Note that we consider parameters $F[k]$ to be located at grid point $k$, so that the localization is applied in the same manner as for $X[k]$. As discussed in Section 2.4.1, we allowed pseudo observations to impact only the associate parameter variables. It means that $L(r)$ is set to 0 if the locations of pseudo observations and analyzed parameters are not identical.

We used root-mean-square error (RMSE) and correlation coefficient (R) as the evaluation metrics. From the 7200 MTUs of the nature run, the simulation of the first 2500 MTUs was discarded as a spin-up of data assimilation. In the remaining 4700 MTUs, we compared the analysis mean of state variables and model parameters with the synthetic truth.

### 3.2. Simple atmospheric general circulation model

To demonstrate the effectiveness of HOOPE-EnKF in addressing realistic problems within Earth system sciences, we employed the SPEEDY model (Molteni 2003). The SPEEDY model is a global atmospheric model with a T30 spectral resolution which is transformed to a Gaussian grid comprising 96 points in the longitudinal direction and 48 points in the latitudinal direction. The model incorporates eight vertical sigma levels. Although the physical parameterizations of the SPEEDY model are more simplistic than those of state-of-the-art models, the SPEEDY model is acknowledged as an effective testbed for examining data assimilation schemes in the context of real atmospheric applications (e.g., Okazaki et al. 2021; Lopez-Restrepo et al. 2021; Sluka et al. 2016; Ruiz et al. 2013a, 2013b; Kang et al. 2011).

We focused on estimating parameters within the convection scheme of the SPEEDY model. This scheme is a simplified version of the mass-flux scheme developed by Tiedke (1993). For a detailed description, please refer to Molteni and Kucharski. The convection scheme is activated in conditionally unstable regions where saturation moist static energy decreases with height and where humidity exceeds a prescribed threshold in both the planetary boundary layer and the layer one level above it. The humidity criteria can be described in the following equation:

$$Q_N > Q_{thr1} \tag{32a}$$
$$Q_{N-1} > Q_{thr2} \tag{32b}$$

where $Q_N$ and $Q_{N-1}$ are specific humidity in the lowest and second lowest vertical layers, respectively. In the original SPEEDY model, the thresholds, $Q_{thr1}$ and $Q_{thr2}$, are defined as functions of saturation specific humidity:

$$Q_{thr1} = RH_{cnv} \times Q_N^{sat} \tag{33a}$$
$$Q_{thr2} = RH_{cnv} \times Q_{N-1}^{sat} \tag{33b}$$

where $Q_N^{sat}$ and $Q_{N-1}^{sat}$ are saturation specific humidity in the lowest and second lowest vertical layers, respectively. $RH_{cnv}$ is a parameter. Ruiz et al. (2013b) found that the model's behavior is notably sensitive to $RH_{cnv}$. They assumed that $RH_{cnv}$ was unknown and varied over time. They successfully estimated this time-varying $RH_{cnv}$ from conventional observations using LETKF.

In the present study, we directly estimated $Q_{thr1}$ and $Q_{thr2}$ without the knowledge that these thresholds are dependent on saturation specific humidity. Given that saturation specific humidity is a function of temperature and pressure, our target parameter is different in the different model grids and its dimension amounts to 9216 (96 grids × 48 grids × 2 levels), in contrast to Ruiz et al. (2013b) who estimated a global parameter. Furthermore, $Q_{thr1}$ and $Q_{thr2}$ temporally change since saturation specific humidity changes. We tested if this spatio-temporally distributed parameter can be estimated under the similar observation settings to Ruiz et al. (2013b).

We performed an observation system simulation experiment. Observations were taken at the grids shown in Figure C1 of Appendix C, spanning every vertical level of the model grid. We observed wind components, temperature, specific humidity, and surface pressure with respective observation errors of 1.0 [ms$^{-1}$], 1.0 [K], 1.0 [gkg$^{-1}$], and 1.0 [hPa]. These observation errors served as the standard deviations of the Gaussian white noise added to the nature run. This setting of observation errors is identical to Ruiz et al. (2013b). For the nature run, we ran the original SPEEDY model with $RH_{cnv} = 0.5$ from January 1982 to March 1983. Observations are collected at 6-hour intervals, matching the frequency of data assimilation cycles.

In the offline batch optimization, we assumed that both $Q_{thr1}$ and $Q_{thr2}$ represent a global time-invariant parameter, denoted as $\widehat{Q_{thr}}$. First, we generated 100 ensemble members of $\widehat{Q_{thr}}$, ranging from $\widehat{Q_{thr}} = 1.0$ to $\widehat{Q_{thr}} = 20.0$ with constant intervals and then executed the SPEEDY model. Note that we no longer applied the equation (33), and the convection was activated when humidity exceeds a fixed threshold. Second, we calculated a climatological index from the ensemble simulations, represented as $\gamma^f$ in Equation 3. This index was derived from the year-averaged specific humidity at the observed grids. An average of 6-hourly simulated specific humidity from January 1982 to December 1982 was taken at the observed grids depicted in Figure C1 and then compared with the averaged observed specific humidity. Note that the dimension of $\gamma^f$ is identical to the number of observable grids and it is much larger than that in the two-scale Lorenz 96 model experiment. Therefore, we decided to construct the surrogate model to directly estimate square differences between simulation and observation. Third, we smoothed the sampled relationship between $\widehat{Q_{thr}}$ and $\gamma^f$ by the Gaussian process regression to construct the surrogate model shown in Equation 4. Fourth, by comparing $\gamma^f$ estimated by the surrogate model with $\gamma^o$, we drew the posterior samples of $\widehat{Q_{thr}}$ by the Metropolis-Hasting algorithm. The observation error of $\gamma^o$ was designated as 1.0 [gkg$^{-1}$] mirroring the original observation error for specific humidity. After discarding the first 100,000 samples as the spin-up period, the remaining 400,000 samples were used to estimate $N(\boldsymbol{\theta}_c, \boldsymbol{C})$. For details on the implementation of the offline batch optimization, see Appendix B.

As we mentioned in section 3.1, we performed three flavors of LETKF: NOHOOPE, HOOPE-EnKF-PSO, and HOOPE-EnKF-RTC. Wind components, temperature, and surface pressure were analyzed every 6 hours. $Q_{thr1}$ and $Q_{thr2}$ were concatenated to the state vector and analyzed every 2 days. We found that frequently updating $Q_{thr1}$ and $Q_{thr2}$ led to unstable estimations. This instability arises because observable state variables are sensitive to $Q_{thr1}$ and $Q_{thr2}$ only when the model and/or the nature run exhibit convective activity at each model grid. Consequently, we updated them less frequently. Note that we refrained from adjusting specific humidity to increase the sensitivity of $Q_{thr1}$ and $Q_{thr2}$ to discrepancies between simulation and observation of specific humidity, ensuring accurate parameter estimation. We adopted the same prior multiplicative inflation that we used in the two-scale Lorenz 96 experiments (see section 3.1) although our focus was solely on experiments with adaptive inflation. The localization function was defined as:

$$L(r,z) = \begin{cases} exp\left(-\frac{1}{2}\left(\left(\frac{r}{\sigma_h}\right)^2 + \left(\frac{z}{\sigma_v}\right)^2\right)\right) & if\ r < 2\sqrt{10/3}\sigma_h\ and\ z < 2\sqrt{10/3}\sigma_v \\ 0 & otherwise \end{cases} \quad (34)$$

where $r$ and $z$ are horizontal and vertical distances between observation and analyzed grid points, respectively. The horizontal and vertical localization scales, $\sigma_h$ and $\sigma_v$, were set to 1000 km and 0.1 (sigma level), respectively. Please refer to section 3.1 for our strategy of localization in the parameter space. The ensemble size was set to 30.

## 4. Results
### 4.1. Two-scale Lorenz 96 model
Figure 2 shows the non-parametric posterior distribution of $\hat{F}$, which can accurately simulate autocorrelations of the system. The dashed red line shows the fitted Gaussian distribution. Although the original sampled distribution is skewed, we directly used sample mean and variance for the Gaussian distribution of the climatological parameter $N(\boldsymbol{\theta}_c, \boldsymbol{C})$.

Figures 3-6 show the performance of three LETKF flavors as a function of the two inflation factors. NOHOOPE has no hope to simultaneously estimate state variables and parameters accurately, even when the ensemble size is extremely large. NOHOOPE can accurately estimate state variables only when it chooses large inflation factors for state variables and small inflation factors for model parameters. This setting of inflation factors prevents the temporal variability of the parameters from being estimated, and makes them converge to a constant value. By specifying large inflation factors for state variables, NOHOOPE can resolve the model error introduced by a constant model parameter. HOOPE-EnKF-PSO and HOOPE-EnKF-RTC substantially stabilize the filters. They accurately estimate both state variables and model parameters within a wide range of inflation factors. This is the empirical evidence that the performance of our proposed method is not greatly sensitive to the choice of covariance inflation. We confirm that they can work even with extremely large inflation factors for model parameters (e.g., 500) (not shown), since when $\rho_\theta \to \infty$, we replace all background samples by their climatological counterparts, ensuring the stable performance of our methods, as discussed in Section 2.4.2. HOOPE-EnKF-RTC is stabler than HOOPE-EnKF-PSO, as we can achieve low RMSE and high R with wider ranges of inflation factors especially when the ensemble size is large. The optimal inflation factors for the state variables are relatively large, representing the large model errors induced by the parameter fixation. Even if it were possible to extensively tune the inflation factors, the performance of NOHOOPE is significantly worse than HOOPE-EnKF-RTC and HOOPE-EnKF-PSO (Tables 1 and 2). Note that it is practically difficult to obtain the optimal performance of NOHOOPE due to the high sensitivity of the performance to inflation factors, while Figures 5 and 6 imply that HOOPE-EnKF-PSO and HOOPE-EnKF-RTC can easily achieve performance similar to the optimal one shown in Tables 1 and 2. It is impractical to extensively tune the inflation factors for maximizing the skill in estimating model parameters since model parameters cannot be observed. Therefore, the insensitivity of the performance to the inflation factors in HOOPE-EnKF-PSO and HOOPE-EnKF-RTC is an important advantage.

To avoid tuning the inflation factors, adaptive inflation was applied. Tables 1 and 2 show that NOHOOPE with adaptive inflation does not work with the small ensemble size. When the ensemble size is large, the adaptive inflation is somehow effective for NOHOOPE to simultaneously estimate state variables and parameters accurately, which was difficult when the inflation factors were fixed. HOOPE-EnKF-PSO and HOOPE-EnKF-RTC outperform NOHOOPE with adaptive inflation. Note that adaptively inflated NOHOOPE with 40 ensemble members does not significantly outperform HOOPE-EnKF-PSO and HOOPE-EnKF-RTC with 10 ensemble members. The results of adaptive inflation indicate that HOOPE-EnKF-RTC is superior to HOOPE-EnKF-PSO due probably to the wider range of inflation factors that yield accurate estimation in HOOPE-EnKF-RTC (Figures 5 and 6). Figures 7 and 8 visualize the estimated $F[k]$ by three LETKF flavors with adaptive inflation. HOOPE-EnKF-PSO and HOOPE-EnKF-RTC track the rapid changes in $F[k]$ more stably and accurately than NOHOOPE. However, the skill of HOOPE-EnKF to estimate extremely small and large values of $F[k]$ is marginal. This is a side effect of introducing climatological parameters to stabilize the filters.

### 4.2. Simple atmospheric general circulation model

Figure 9 shows the non-parametric posterior distribution of $\widehat{Q_{thr}}$, which can accurately simulate the climatology of specific humidity. The dashed red line shows the fitted Gaussian distribution. Notably, the original sampled distribution exhibits significant skewness. The sensitivity of the global distribution of specific humidity to $\widehat{Q_{thr}}$ becomes negligible when $\widehat{Q_{thr}}$ is larger than 15 [gkg$^{-1}$] since convection rarely occurs in this range. The pronounced skewness of the distribution arises from this nonlinear behavior of the parameter's sensitivity. As we did in section 4.1, we directly used sample mean and variance for the Gaussian distribution of the climatological parameter $N(\boldsymbol{\theta}_c, \boldsymbol{C})$.

To assess three LETKFs in estimating state variables, we used specific humidity at the lowest level. No notable differences were found in the other state variables (not shown). Figure 10a shows RMSE between analysis and synthetic true specific humidity. While two variants of HOOPE-EnKF stably maintain RMSE below 1.0 [gkg$^{-1}$], NOHOOPE occasionally gets higher RMSE after March 1982.

This poor performance of NOHOOPE in estimating low level specific humidity can be attributed to its inaccurate estimation of $Q_{thr1}$ and $Q_{thr2}$. Figure 10b shows RMSE between analysis and synthetic true $Q_{thr1}$. The results for $Q_{thr2}$ closely mirror those of $Q_{thr1}$ (not shown). After March 1982, the parameter estimation of NOHOOPE gets unstable, leading to extremely large RMSE. Conversely, both HOOPE-EnKF-PSO and HOOPE-EnKF-RTC offer accurate and consistent estimations of $Q_{thr1}$. Figure 11 shows the global distribution of $RH_{cnv} \times Q_N^{sat}$ (synthetic true convection thresholds) and estimated $Q_{thr1}$. Both HOOPE-EnKF-PSO and HOOPE-EnKF-RTC reproduce high $Q_{thr1}$ along the intertropical convergence zone. In contrast, NOHOOPE produces extremely large positive and even negative $Q_{thr1}$. Figure 12 shows that both HOOPE-EnKF-PSO and HOOPE-EnKF-RTC accurately estimate the meridional shift of $Q_{thr1}$. Overall, the difference of accuracies between HOOPE-EnKF-PSO and HOOPE-EnKF-RTC is negligible. Their estimated $Q_{thr1}$ in mid-latitude and polar regions is consistently greater than the synthetic truth. Given that the error in specific humidity in the tropics is significantly more pronounced than in other regions, offline batch optimization tends to prioritize parameters that minimize the error in the tropics, overlooking the comparatively minor errors elsewhere. Consequently, our estimated posterior distribution of $\widehat{Q_{thr}}$ presents considerably higher values than the synthetic true $Q_{thr1}$ in mid-latitude and polar regions. HOOPE-EnKF thus inhibits $Q_{thr1}$ from reaching such low values suitable for mid-latitude and polar regions. This is the side effect of introducing global climatological parameters to stabilize the filters.

## 5. Discussion and Conclusions

We proposed combining offline batch optimization and EnKF toward an efficient and practical method to estimate time-varying parameters in high-dimensional spaces. In our newly proposed method, HOOPE-EnKF, model parameters estimated by EnKF are constrained by the results of offline batch optimization, in which the posterior distribution of model parameters is obtained by comparing simulated and observed climatological variables. HOOPE-EnKF outperforms the original LETKF in the synthetic experiments using the two-scale Lorenz96 model and the SPEEDY atmospheric model. One advantage of HOOPE-EnKF over the traditional EnKFs is that its performance is not greatly affected by inflation factors for model parameters, thus eliminating the need for extensive tuning of inflation factors. The inflation of the online estimated error of model parameters has been extensively discussed in previous works (e.g., Kotsuki et al. 2018; Ruiz et al. 2013a; Moradkhani et al. 2005b), and HOOPE-EnKF contributes to this body of published literature.

Despite numerous studies related to EnKFs that address the estimation of high-dimensional parameters, including real-world weather forecasting (e.g., Ruckstuhl and Janjić 2020), our work offers a unique contribution to this established literature. Pulido et al. (2016) suggested that sub-grid scale contributions of the two-scale Lorenz 96 model were initially parameterized by variables at a larger scale. Then, a functional form of this parameterization was estimated online by ETKF. While the approach by Pulido et al. (2016) successfully reduced the dimension of targeted parameters, the advantage of HOOPE-EnKF over this method lies in the absence of a need for a priori knowledge of the functional form. Instead, we directly estimate high-dimensional parameters. Kang et al. (2011) and Bellsky et al. (2014) emphasized the significance of localizing the optimization problem of high-dimensional parameters to accurately estimate spatio-temporally distributed parameters. However, Kang et al. (2011) could not achieve sufficient skill when they assumed that their targeted parameters varied over time. Although Bellsky et al. (2014) successfully solved a problem similar to ours in the two-scale Lorenz 96 model, they posited two constraints: all model grids had to be observed, and localization scales needed to be narrowly set. It implies a necessity to avoid using remote observations. In contrast, our approach is not bound by these constraints. When remote observations are insensitive to parameters, the parameters are forced to align with climatology, ensuring the stability of the filter. Therefore, we can take a risk to use remote observations to constrain parameters, and it is unnecessary for us to observe all model grids. Katzfuss et al. (2020) rigorously derived new filters and smoothers in which high-dimensional state variables were estimated by EnKF, while parameters were separately estimated by non-parametric Bayesian inference techniques such as Gibbs sampling and particle filtering. Although their method holds promise for joint state-parameter estimation, it is best suited for scenarios where the number of unknown parameters is not too large, which differs from the context of our study.

We demonstrated that the estimation of spatio-temporally distributed model parameters using HOOPE-EnKF can contribute to the quantification of systematic model errors introduced by unresolved physics. Data-driven quantification and reduction of systematic model errors within data assimilation frameworks have been actively investigated (e.g., Pathiraja and van Leeuwen 2022; Amemiya et al. 2023; Tomizawa and Sawada 2021; Brajard et al. 2020; Dueben and Bauer 2018; Weyn et al. 2019, 2021), and HOOPE-EnKF can contribute to this body of existing literature. Given the recent growth in the applications of efficient uncertainty quantification methods to components of Earth system models (e.g., Dunbar et al. 2021; Zhang et al. 2020; Sawada 2020; Teixeira Parente et al. 2019; Duan et al. 2017), the additional computational cost for offline batch optimization may be comparable to the training costs of the existing data-driven approaches cited above.

Compared to other methods cited above, our proposed method can be more readily applied to numerical weather prediction. After obtaining the climatology of model parameters by offline batch optimization, we can directly use the existing architecture of EnKF and a process-based model, which significantly differs from the non-traditional prediction styles employed by neural networks to mitigate biases in process-based models (e.g., Tomizawa and Sawada 2021; Brajard et al. 2020; Dueben and Bauer 2018; Weyn et al. 2019; 2021). HOOPE-EnKF does not require extremely accurate observations, while Pathiraja and van Leeuwen (2022) necessitated observations with minimal errors. In HOOPE-EnKF, the entire system does not need to be observed, whereas the observation network of Amemiya et al. (2023) included observations in all grid points. HOOPE-EnKF does not rely on the assumption of homogeneity in sub-grid scale physics, which is a requirement for Pathiraja and van Leeuwen (2022) and Amemiya et al. (2023). When a model parameter is not sensitive to any observations assimilated within a data assimilation window, HOOPE-EnKF ensures that this model parameter converges to $N(\boldsymbol{\theta}_c, \boldsymbol{C})$. This property of HOOPE-EnKF may significantly contribute to the stability of the filter avoiding the negative impact of spurious correlation in numerical weather prediction, where the sensitivity of observation to model parameters varies spatio-temporally and the observation network is highly heterogeneous. It is worth noting that our HOOPE-EnKF concept can be extended to variational methods such as 4D-VAR, commonly used in the operational weather prediction systems, because our proposed cost function (Equation 12) and method to solve it can be shared with variational methods.

The primary disadvantage of HOOPE-EnKF is its inability to predict systematic model errors, while Pathiraja and van Leeuwen (2022) and Amemiya et al. (2023) successfully constructed the statistical models to directly predict the error term induced by unresolved processes. Although HOOPE-EnKF accurately estimates the contribution of unresolved physics, it still assumes a persistent model for the evolution of model parameters during the forecast phase. Consequently, the prediction error may increase for lead times longer than the timescale of change in time-varying parameters. A reasonable countermeasure to this limitation is to nudge the time-varying parameters towards $N(\boldsymbol{\theta}_c, \boldsymbol{C})$ during a forecast step. It is promising to analyze the spatio-temporal distribution of parameters in specific parameterizations and deepen the understanding of their uncertainty, as demonstrated by our SPEEDY model experiment. By understanding the source of uncertainty in a specific parameterization from spatio-temporally distributed model parameters, it becomes straightforward to predict and reduce the systematic model error.

We proposed two variants of HOOPE-EnKF: HOOPE-EnKF-PSO and HOOPE-EnKF-RTC. Both variants perform significantly better than the vanilla LETKF, and their performances are less sensitive to inflation factors. Apparently, the advantage of HOOPE-EnKF-PSO is its ease of implementation. HOOPE-EnKF-PSO only requires generating pseudo-observations of model parameters to be assimilated and no modification of a data assimilation code is necessary. Although HOOPE-EnKF-RTC is more complex than HOOPE-EnKF-PSO, the additional computational cost is comparable to HOOPE-EnKF-PSO in our proposed implementation. HOOPE-EnKF-RTC is better than HOOPE-EnKF-PSO in our experiment of the two-scale Lorenz96 model. This is likely because, in HOOPE-EnKF-RTC, we conditioned ensemble members of model parameters before assimilating observations, which contribute to eliminating spurious correlations between observation and model parameters. However, no notable differences are found in the more realistic experiment by the simple global atmospheric SPEEDY model. Our future work should focus on testing both methods in real-world applications, such as numerical weather prediction.

**Acknowledgements**

This work was supported by JST Moonshot R&D program (grant no. JPMJMS2281). We used the Wisteria-BDEC supercomputer at the University of Tokyo under Joint Usage/Research Center for Interdisciplinary Large-scale Information Infrastructures (JHPCN) in Japan (Project ID: jh220020, jh230003).

**Open Research**

The source code of this work can be found at https://doi.org/10.5281/zenodo.8400926 (Sawada, 2023). Note that the original source code of LETKF is available at Github (Miyoshi, 2016 https://github.com/takemasa-miyoshi/letkf) and this code was adopted in our two-scale Lorenz96 model experiment. In our SPEEDY atmospheric model experiment, we adopted the code archived in https://zenodo.org/record/1198432 (Hatfield, 2018).

**Appendix**

**A. Analytic form of the sum of the background and climatological terms**

Our problem is to combine the sum of the background and climatological terms in (12) into a general quadratic form:

$$J = \frac{1}{2}\begin{pmatrix}x-\bar{x}^b\\\theta-\bar{\theta}^b\end{pmatrix}^T \begin{pmatrix}B_x & B_{x\theta}\\B_{\theta x} & B_\theta\end{pmatrix}^{-1}\begin{pmatrix}x-\bar{x}^b\\\theta-\bar{\theta}^b\end{pmatrix} + \frac{1}{2}(\theta-\theta_c)^T C^{-1}(\theta-\theta_c)$$

$$= \frac{1}{2}\begin{pmatrix}x-\tilde{x}^b\\\theta-\tilde{\theta}^b\end{pmatrix}^T \begin{pmatrix}\tilde{B}_x & \tilde{B}_{x\theta}\\\tilde{B}_{\theta x} & \tilde{B}_\theta\end{pmatrix}^{-1}\begin{pmatrix}x-\tilde{x}^b\\\theta-\tilde{\theta}^b\end{pmatrix} + const \quad (A1)$$

To find the exact forms of the tilde variables on the right-hand side of (A1), we rely on the fact that the product of two Gaussian pdfs is also a Gaussian pdf up to a multiplicative factor (see Bishop (2006) for the proof). However, a difficulty here is that the climatology term does not contain any term related to $x$, i.e., a Gaussian pdf in a degenerate form. Therefore, we will apply a limitation process by introducing a pseudo climatological error covariance $D = \delta I$ and an arbitrarily climatological mean $x_c$ for $x$, and let $\delta$ go to infinity to get the result.

$$J = \frac{1}{2}\begin{pmatrix}x-\bar{x}^b\\\theta-\bar{\theta}^b\end{pmatrix}^T \begin{pmatrix}B_x & B_{x\theta}\\B_{\theta x} & B_\theta\end{pmatrix}^{-1}\begin{pmatrix}x-\bar{x}^b\\\theta-\bar{\theta}^b\end{pmatrix} + \frac{1}{2}\begin{pmatrix}x-x_c\\\theta-\theta_c\end{pmatrix}^T \begin{pmatrix}D & 0\\0 & C\end{pmatrix}^{-1}\begin{pmatrix}x-x_c\\\theta-\theta_c\end{pmatrix}$$

$$= \frac{1}{2}\begin{pmatrix}x-\tilde{x}^b\\\theta-\tilde{\theta}^b\end{pmatrix}^T \begin{pmatrix}\tilde{B}_x & \tilde{B}_{x\theta}\\\tilde{B}_{\theta x} & \tilde{B}_\theta\end{pmatrix}^{-1}\begin{pmatrix}x-\tilde{x}^b\\\theta-\tilde{\theta}^b\end{pmatrix} + const \quad (A2)$$

The analytical forms of the tilde variables can be found from the two following equations (see Bishop (2006))

$$\begin{pmatrix}\tilde{B}_x & \tilde{B}_{x\theta}\\\tilde{B}_{\theta x} & \tilde{B}_\theta\end{pmatrix}^{-1} = \begin{pmatrix}B_x & B_{x\theta}\\B_{\theta x} & B_\theta\end{pmatrix}^{-1} + \begin{pmatrix}D & 0\\0 & C\end{pmatrix}^{-1} \quad (A3)$$

$$\begin{pmatrix}\tilde{x}^b\\\tilde{\theta}^b\end{pmatrix} = \begin{pmatrix}\tilde{B}_x & \tilde{B}_{x\theta}\\\tilde{B}_{\theta x} & \tilde{B}_\theta\end{pmatrix}\begin{pmatrix}B_x & B_{x\theta}\\B_{\theta x} & B_\theta\end{pmatrix}^{-1}\begin{pmatrix}\bar{x}^b\\\bar{\theta}^b\end{pmatrix} + \begin{pmatrix}\tilde{B}_x & \tilde{B}_{x\theta}\\\tilde{B}_{\theta x} & \tilde{B}_\theta\end{pmatrix}\begin{pmatrix}D & 0\\0 & C\end{pmatrix}^{-1}\begin{pmatrix}x_c\\\theta_c\end{pmatrix} \quad (A4)$$

We first use a matrix identity derived from the Woodbury matrix identity for any symmetric positive-definite matrices $A, B$

$$[A^{-1}+B^{-1}]^{-1} = A - A[A+B]^{-1}A \quad (A5)$$

to rewrite (A3)

$$\begin{pmatrix}\tilde{B}_x & \tilde{B}_{x\theta}\\\tilde{B}_{\theta x} & \tilde{B}_\theta\end{pmatrix} = \begin{pmatrix}D & 0\\0 & C\end{pmatrix} - \begin{pmatrix}D & 0\\0 & C\end{pmatrix}\left[\begin{pmatrix}D & 0\\0 & C\end{pmatrix} + \begin{pmatrix}B_x & B_{x\theta}\\B_{\theta x} & B_\theta\end{pmatrix}\right]^{-1}\begin{pmatrix}D & 0\\0 & C\end{pmatrix} \quad (A6)$$

where we have replaced $A = \begin{pmatrix}D & 0\\0 & C\end{pmatrix}$ and $B = \begin{pmatrix}B_x & B_{x\theta}\\B_{\theta x} & B_\theta\end{pmatrix}$. Our remaining task is to find the inverse of the matrix in the second term of the right-hand side of (A6)

$$\begin{pmatrix} E & F \\ G & H \end{pmatrix} = \begin{pmatrix} D + B_x & B_{x\theta} \\ B_{\theta x} & C + B_\theta \end{pmatrix} \tag{A7}$$

The block-wise analytic inverse formula gives the following result

$$\begin{pmatrix} \widetilde{E} & \widetilde{F} \\ \widetilde{G} & \widetilde{H} \end{pmatrix} = \begin{pmatrix} E & F \\ G & H \end{pmatrix}^{-1} = \begin{pmatrix} E^{-1} + E^{-1}FS^{-1}GE^{-1} & -E^{-1}FS^{-1} \\ -S^{-1}GE^{-1} & S^{-1} \end{pmatrix} \tag{A8}$$

where $S = H - GE^{-1}F$ is the Schur complement of $E$.

From (A6), (A7) and (A8) we can now derive the analytic forms of the tilde matrices in (A3)

$$\begin{pmatrix} \widetilde{B}_x & \widetilde{B}_{x\theta} \\ \widetilde{B}_{\theta x} & \widetilde{B}_\theta \end{pmatrix} = \begin{pmatrix} D - D\widetilde{E}D & -D\widetilde{F}C \\ -C\widetilde{G}D & C - C\widetilde{H}C \end{pmatrix} \tag{A9}$$

First it is easy to obtain the formula of $\widetilde{B}_\theta$

$$\widetilde{B}_\theta = C - C[C + B_\theta - B_{\theta x}[D + B_x]^{-1}B_{x\theta}]^{-1}C \tag{A10}$$

When $D \to \infty$, (A10) becomes

$$\widetilde{B}_\theta = C - C[C + B_\theta]^{-1}C \tag{A11}$$

The identity (A5) points out that

$$\widetilde{B}_\theta^{-1} = B_\theta^{-1} + C^{-1} \tag{A12}$$

In other words, $\widetilde{B}_\theta$ can be derived directly from the marginal prior pdf of $\theta$ and its climatological pdf.

Since $\widetilde{B}_{\theta x}$ is the transpose of $\widetilde{B}_{x\theta}$, we only need to write the formula of $\widetilde{B}_{x\theta}$ here

$$\widetilde{B}_{x\theta} = D[D + B_x]^{-1}B_{x\theta}[C + B_\theta - B_{\theta x}[D + B_x]^{-1}B_{x\theta}]^{-1}C \tag{A13}$$

Note that $D[D + B_x]^{-1} = [I + B_x D^{-1}]^{-1}$, when $D \to \infty$, (A13) becomes

$$\widetilde{B}_{x\theta} = B_{x\theta}[C + B_\theta]^{-1}C \tag{A14}$$

When $C, B_\theta$ are diagonal matrices, (A14) shows that new correlations between $x$ and $\theta$ are slightly reduced by introducing the climatological constraint on $\theta$.

Finally, $\widetilde{B}_x$ has the most complicated form

$$\widetilde{B}_x = D - D[D + B_x]^{-1}D$$
$$- D[D + B_x]^{-1}B_{x\theta}[C + B_\theta - B_{\theta x}[D + B_x]^{-1}B_{x\theta}]^{-1}B_{\theta x}[D + B_x]^{-1}D \tag{A15}$$

Clearly, it is difficult to take limitation on such a complicated form. A better strategy is exchanging the roles of $x$ and $\theta$ to obtain a simpler form similar to (A10)

$$\widetilde{B}_x = D - D[D + B_x - B_{x\theta}[C + B_\theta]^{-1}B_{\theta x}]^{-1}D \tag{A16}$$

We again use (A5) to rewrite this identity as

$$\widetilde{B}_x^{-1} = D^{-1} + [B_x - B_{x\theta}[C + B_\theta]^{-1}B_{\theta x}]^{-1} \tag{A17}$$

Then it is easy to find that when $D \to \infty$, we have

$$\widetilde{B}_x = B_x - B_{x\theta}[C + B_\theta]^{-1}B_{\theta x} \qquad (A18)$$

From (A12), (A14) and (A18), we can write explicitly the analytic form of the updated background error covariance

$$\begin{pmatrix} \widetilde{B}_x & \widetilde{B}_{x\theta} \\ \widetilde{B}_{\theta x} & \widetilde{B}_\theta \end{pmatrix} = \begin{pmatrix} B_x - B_{x\theta}[C + B_\theta]^{-1}B_{\theta x} & B_{x\theta}[C + B_\theta]^{-1}C \\ C[C + B_\theta]^{-1}B_{\theta x} & [C^{-1} + B_\theta^{-1}]^{-1} \end{pmatrix} \qquad (A19)$$

Plugging (A19) into (A4), the updated first moments can be derived now

$$\begin{pmatrix} \widetilde{x}^b \\ \widetilde{\theta}^b \end{pmatrix} = \begin{pmatrix} \widetilde{B}_x & \widetilde{B}_{x\theta} \\ \widetilde{B}_{\theta x} & \widetilde{B}_\theta \end{pmatrix} \begin{pmatrix} B_x & B_{x\theta} \\ B_{\theta x} & B_\theta \end{pmatrix}^{-1} \begin{pmatrix} \overline{x}^b \\ \overline{\theta}^b \end{pmatrix} + \begin{pmatrix} \widetilde{B}_{x\theta} \\ \widetilde{B}_\theta \end{pmatrix} C^{-1}\theta_c \qquad (A20)$$

After some mathematical operations using the block-wise inverse identity (A8), the following analytic formula is obtained

$$\begin{pmatrix} \widetilde{x}^b \\ \widetilde{\theta}^b \end{pmatrix} = \begin{pmatrix} \overline{x}^b + B_{x\theta}[C + B_\theta]^{-1}(\theta_c - \overline{\theta}^b) \\ C[C + B_\theta]^{-1}\theta_b + B_\theta[C + B_\theta]^{-1}\theta_c \end{pmatrix} \qquad (A21)$$

Like $\widetilde{B}_\theta$, this states that $\widetilde{\theta}^b$ can also be derived directly from the marginal prior pdf of $\theta$ and its climatological pdf

$$\widetilde{\theta}^b = \widetilde{B}_\theta B_\theta^{-1}\overline{\theta}^b + \widetilde{B}_\theta C^{-1}\theta_c \qquad (A22)$$

We check the validity of (A19) and (A20) under some special cases. Clearly, when the climatological constraint is not applied, i.e., $C \to \infty$, both reduce to the original moments. It is more interesting to check the reverse case when $C \to 0$, i.e., the parameters are fixed as $\theta_c$

$$\begin{pmatrix} \widetilde{B}_x & \widetilde{B}_{x\theta} \\ \widetilde{B}_{\theta x} & \widetilde{B}_\theta \end{pmatrix} = \begin{pmatrix} B_x - B_{x\theta}B_\theta^{-1}B_{\theta x} & 0 \\ 0 & 0 \end{pmatrix} \qquad (A23)$$

$$\begin{pmatrix} \widetilde{x}^b \\ \widetilde{\theta}^b \end{pmatrix} = \begin{pmatrix} \overline{x}^b + B_{x\theta}B_\theta^{-1}(\theta_c - \overline{\theta}^b) \\ \theta_c \end{pmatrix} \qquad (A24)$$

The updated $\widetilde{x}^b$ and $\widetilde{B}_x$ are exactly the first and second moments, respectively, of the conditional pdf of $x$ on $\theta$, which is also the Schur complement of $B_x$. Finally, when $x$ and $\theta$ are uncorrelated, we have

$$\begin{pmatrix} \widetilde{B}_x & \widetilde{B}_{x\theta} \\ \widetilde{B}_{\theta x} & \widetilde{B}_\theta \end{pmatrix} = \begin{pmatrix} B_x & 0 \\ 0 & [C^{-1} + B_\theta^{-1}]^{-1} \end{pmatrix} \qquad (A25)$$

$$\begin{pmatrix} \widetilde{x}^b \\ \widetilde{\theta}^b \end{pmatrix} = \begin{pmatrix} \overline{x}^b \\ C[C + B_\theta]^{-1}\overline{\theta}^b + B_\theta[C + B_\theta]^{-1}\theta_c \end{pmatrix} \qquad (A26)$$

As expected, this reflects that only $\theta$ should be updated in this case.

## B. Detailed description of offline batch optimization

The Metropolis-Hastings algorithm (Hastings 1970) was used as a MCMC sampler in offline batch optimization. The implementation of the MCMC sampler is the following:

*1. For each iteration j, generate a candidate parameter vector $\widehat{\boldsymbol{\theta}}_{candidate}$.*
This candidate is sampled from the proposal distribution $q(\widehat{\boldsymbol{\theta}}_{candidate}|\widehat{\boldsymbol{\theta}}_j)$.

*2. Calculating an acceptance probability of $\widehat{\boldsymbol{\theta}}_{candidate}$, $\alpha(\widehat{\boldsymbol{\theta}}_j, \widehat{\boldsymbol{\theta}}_{candidate})$:*

$$\alpha(\widehat{\boldsymbol{\theta}}_j, \widehat{\boldsymbol{\theta}}_{candidate}) = \exp\left(\Phi^{(s)}(\widehat{\boldsymbol{\theta}}_j) - \Phi^{(s)}(\widehat{\boldsymbol{\theta}}_{candidate})\right) \quad (B1)$$

where $\Phi^{(s)}(\widehat{\boldsymbol{\theta}})$ is a likelihood function. In the experiment of the two-scale Lorenz 96 model, the functional form of $\Phi^{(s)}(\widehat{\boldsymbol{\theta}})$ is the following:

$$\Phi^{(s)}(\widehat{\boldsymbol{\theta}}) = \frac{1}{2} \frac{\|\boldsymbol{\gamma}^o - g^{(s)}(\widehat{\boldsymbol{\theta}})\|^2}{R_{gp}(\widehat{\boldsymbol{\theta}}) + R_o} \quad (B2)$$

where $R_{gp}(\widehat{\boldsymbol{\theta}})$ and $R_o$ are the error variances of $g^{(s)}(\widehat{\boldsymbol{\theta}})$ (i.e. errors in a surrogate model calculated by Gaussian process) and observation, respectively. In the experiment of the SPEEDY model, this square difference is directly approximated by Gaussian process regression, so that the functional form of $\Phi^{(s)}(\widehat{\boldsymbol{\theta}})$ is:

$$\Phi^{(s)}(\widehat{\boldsymbol{\theta}}) = \frac{1}{2} \frac{g^{(s)}(\widehat{\boldsymbol{\theta}})}{R_{gp}(\widehat{\boldsymbol{\theta}}) + R_o} \quad (B3)$$

See also Equations (4a) and (4b). The acceptance probability is calculated based on the square difference between the simulated and observed climatological indices normalized by the total error variance of the Gaussian process and the observation measurement.

*3. Determine if $\widehat{\boldsymbol{\theta}}_{candidate}$ is accepted as a new parameter or not.*
A random number, *b*, is generated from the uniform distribution of [0,1]. Then,
If $b \leq \alpha(\widehat{\boldsymbol{\theta}}_j, \widehat{\boldsymbol{\theta}}_{candidate})$, accept the candidate parameter and $\widehat{\boldsymbol{\theta}}_{j+1} = \widehat{\boldsymbol{\theta}}_{candidate}$
If $b > \alpha(\widehat{\boldsymbol{\theta}}_j, \widehat{\boldsymbol{\theta}}_{candidate})$, reject the candidate parameter and $\widehat{\boldsymbol{\theta}}_{j+1} = \widehat{\boldsymbol{\theta}}_j$

These three steps were iterated 500,000 times in this study and the first 100,000 iterations were discarded as the spin-up period. From the remaining 400,000 samples, the probabilistic distribution of parameters was obtained. In this study, $q(\widehat{\boldsymbol{\theta}}_{candidate}|\widehat{\boldsymbol{\theta}}_j)$ is assumed to be Gaussian with zero mean. Since the Gaussian process can provide both mean and variance of their estimation in each point, it is straightforward to obtain $R_{gp}(\widehat{\boldsymbol{\theta}})$ whenever $g^{(s)}(\widehat{\boldsymbol{\theta}})$ is evaluated. To calculate $R_o$, the subset of the continuous timeseries of observations was randomly chosen and 1,000 observed climatological indices $\boldsymbol{\gamma}^o$ (i.e. autocorrelation) were generated in the two-scale Lorenz96 model experiment. The variance of these observed climatological indices was recognized as $R_o$. In the SPEEDY model experiment, we directly used the prescribed observation error in specific humidity as $R_o$.

**C. Observation network in the SPEEDY model experiment**

The simulated observation network in the SPEEDY model experiment can be found in Figure C1.

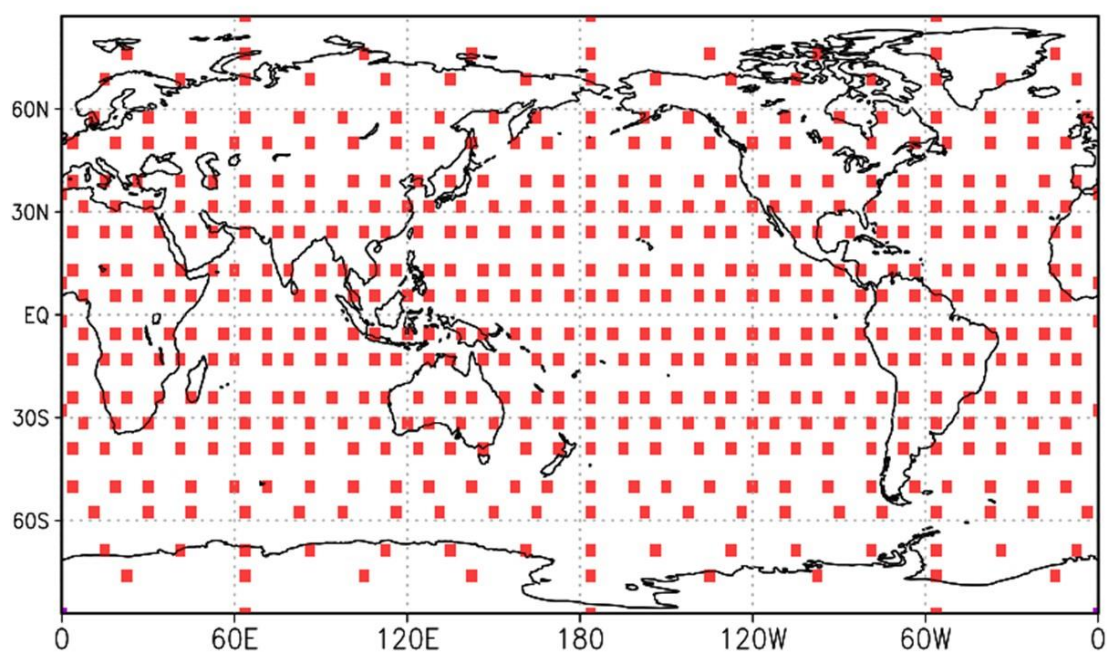

**Figure C1**. Red squares show the location of the simulated observations in the SPEEDY model experiment.

**Table 1.** Summary of RMSE of model parameters in three LETKF variants. "Optimal" means the performance by inflation factors which minimize RMSE.

| ensemble size | NOHOOPE | | HOOPE-EnKF-PSO | | HOOPE-EnKF-RTC | |
|---:|---:|---:|---:|---:|---:|---:|
| | Optimal | Adaptive | Optimal | Adaptive | Optimal | Adaptive |
| 10 | 4.475 | N/A | 2.547 | 2.785 | 2.498 | 2.736 |
| 20 | 2.906 | 3.910 | 2.169 | 2.313 | 2.174 | 2.265 |
| 40 | 2.955 | 3.716 | 2.159 | 2.304 | 2.162 | 2.254 |

**Table 2.** Summary of R of model parameters in three LETKF variants. "Optimal" means the performance by inflation factors which maximize R.

| ensemble size | NOHOOPE | | HOOPE-EnKF-PSO | | HOOPE-EnKF-RTC | |
|---:|---:|---:|---:|---:|---:|---:|
| | Optimal | Adaptive | Optimal | Adaptive | Optimal | Adaptive |
| 10 | 0.029 | N/A | 0.366 | 0.260 | 0.336 | 0.261 |
| 20 | 0.297 | 0.255 | 0.535 | 0.421 | 0.531 | 0.458 |
| 40 | 0.306 | 0.268 | 0.541 | 0.424 | 0.539 | 0.463 |

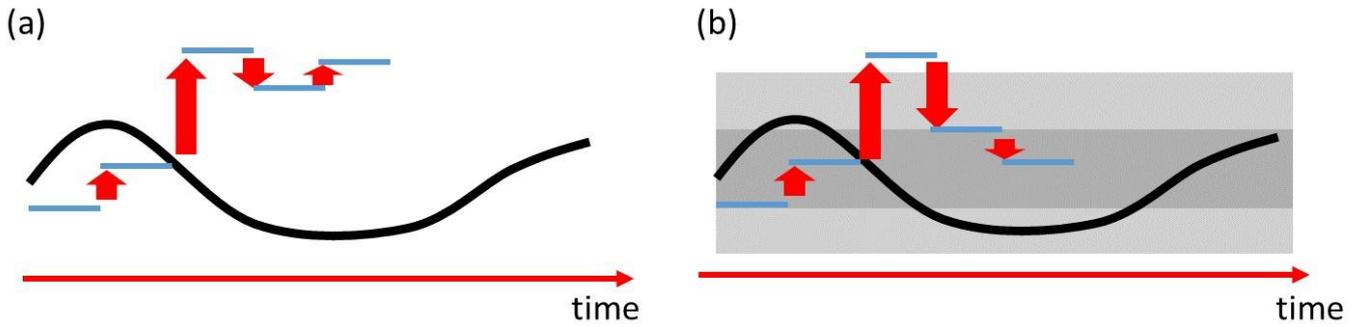

**Figure 1.** Schematics of the (a) original sequential data assimilation and (b) Hybrid Online and Offline Parameter Estimation (HOOPE) concept. Black lines are timeseries of the true model parameter. Blue lines are the estimation of the online data assimilation algorithm and red arrows show the adjustments by the data assimilation steps. The gray area in (b) shows the posterior distribution of model parameters calculated by the offline batch optimization. When estimated parameters substantially deviate from the truth due to erroneous observations and/or sampling error with an insufficient ensemble size, there is a risk of filter degeneracy. HOOPE-EnKF is expected to prevent filter degeneracy in this situation by efficiently move ensemble members to the probabilistic distribution of time-invariant parameters. See also Section 2. This figure is adopted from Sawada (2022).

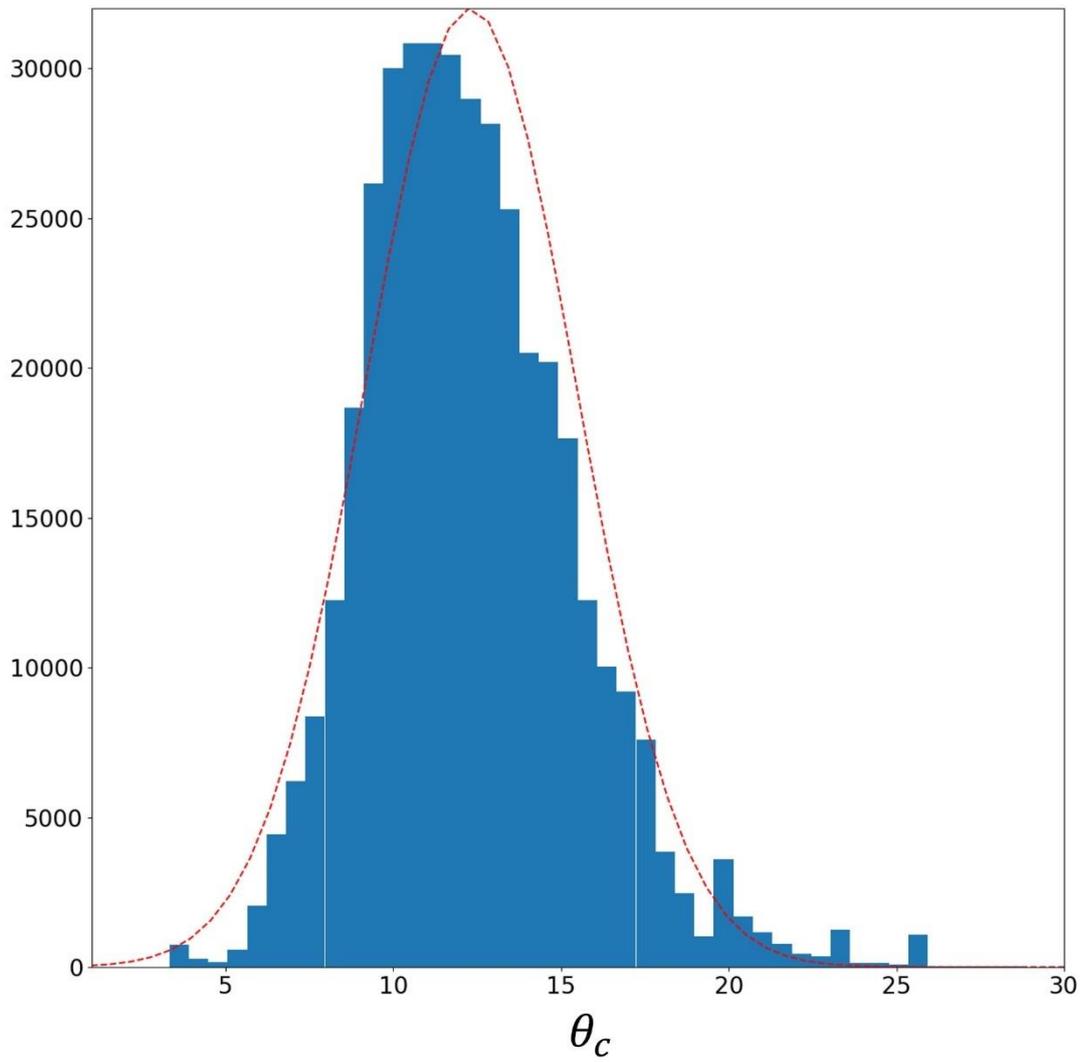

**Figure 2.** Posterior distribution of a time-invariant model parameter from (blue) samples by the Metropolis-Hastings algorithm and (red) a fitted Gaussian distribution in the two-scale Lorenz 96 experiment.

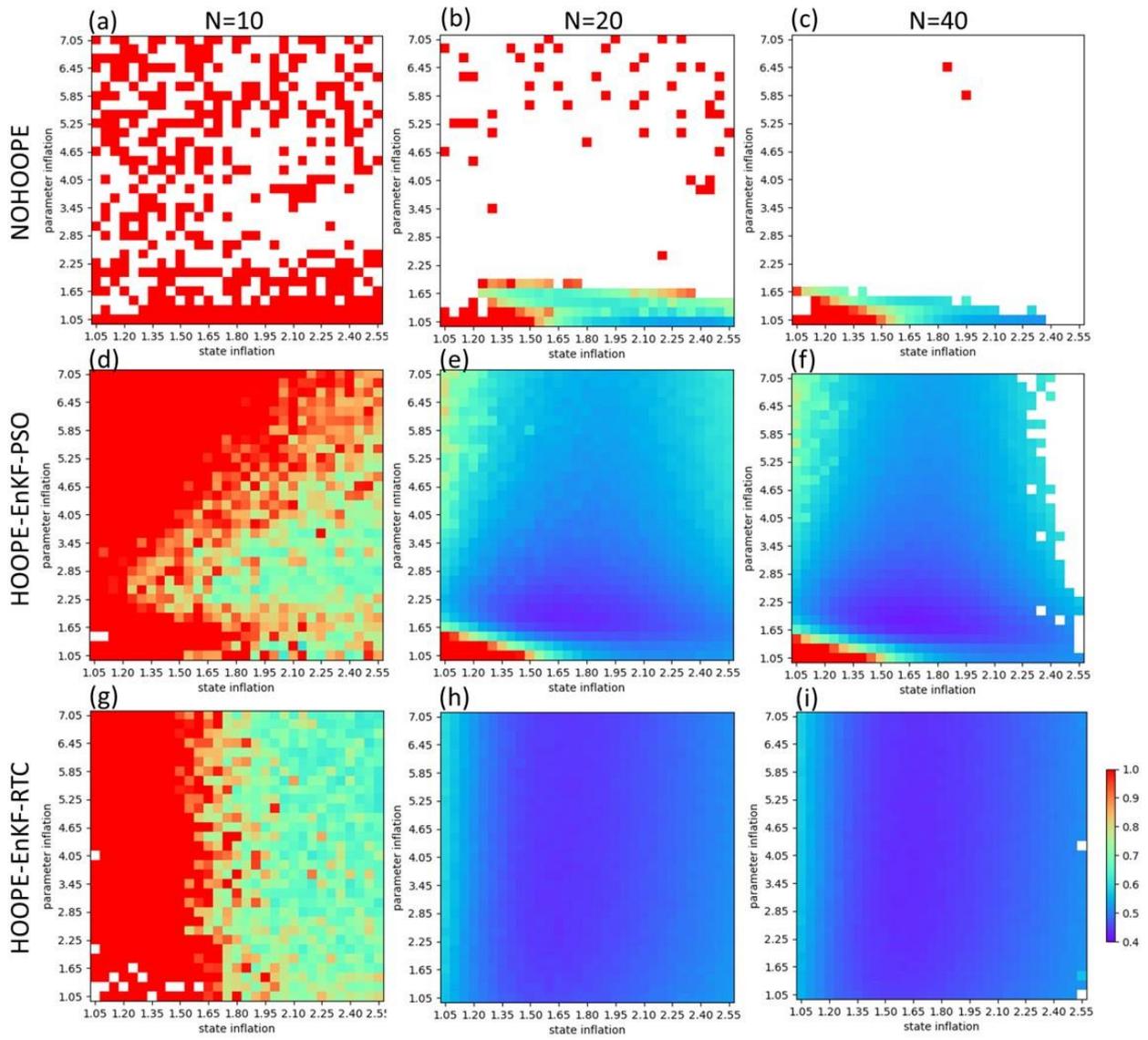

**Figure 3**. RMSE of state variables in (a-c) NOHOOPE, (d-f) HOOPE-EnKF-PSO, and (g-i) HOOPE-EnKF-RTC with the ensemble size of (a,d,g) 10, (b,e,h) 20, and (c,f,i) 40. Horizontal axis shows inflation factors for state variables, while vertical axis shows inflation factors for model parameters. White tiles indicate that the model diverges.

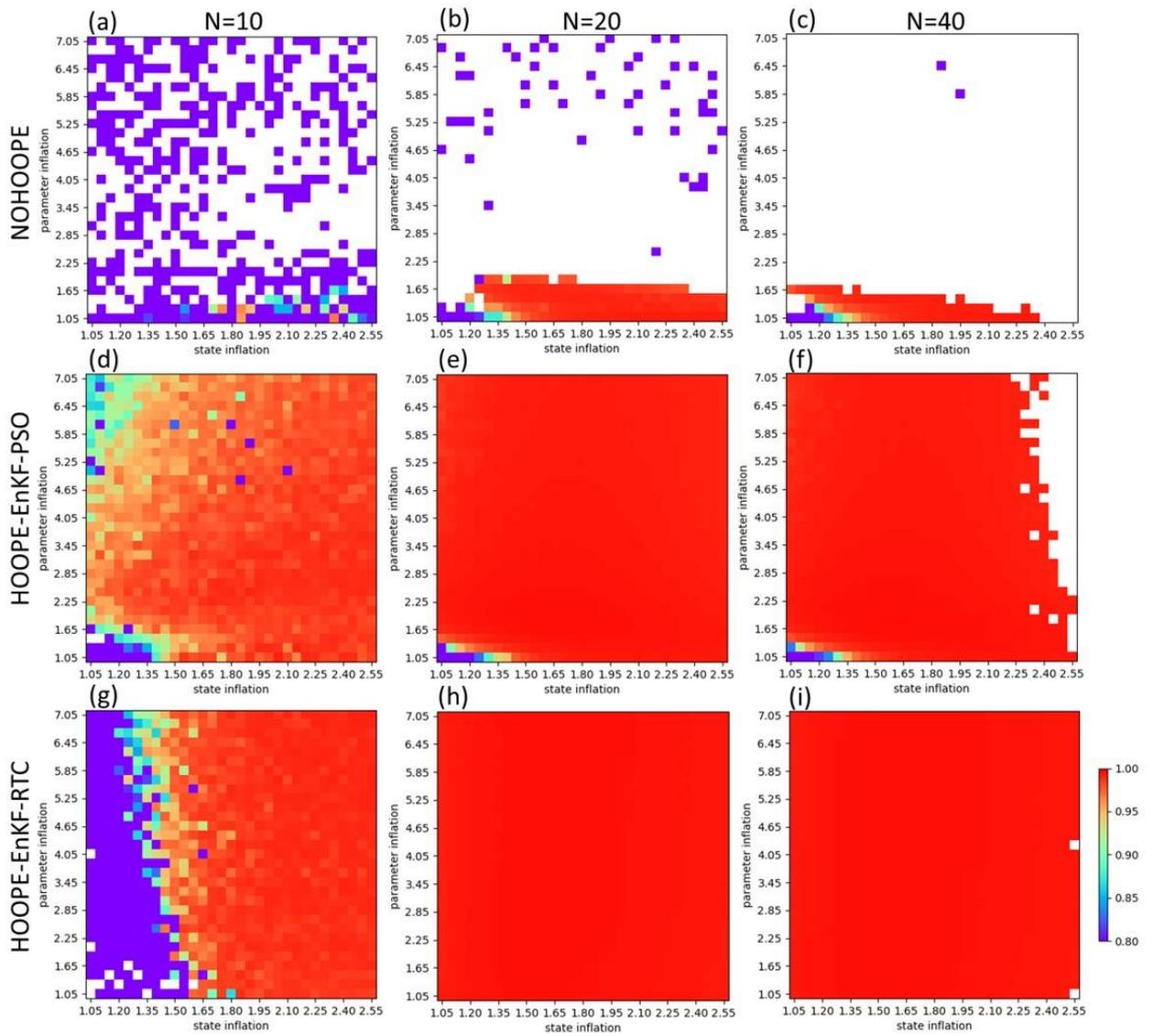

**Figure 4**. Same as Figure 3 but for R of state variables.

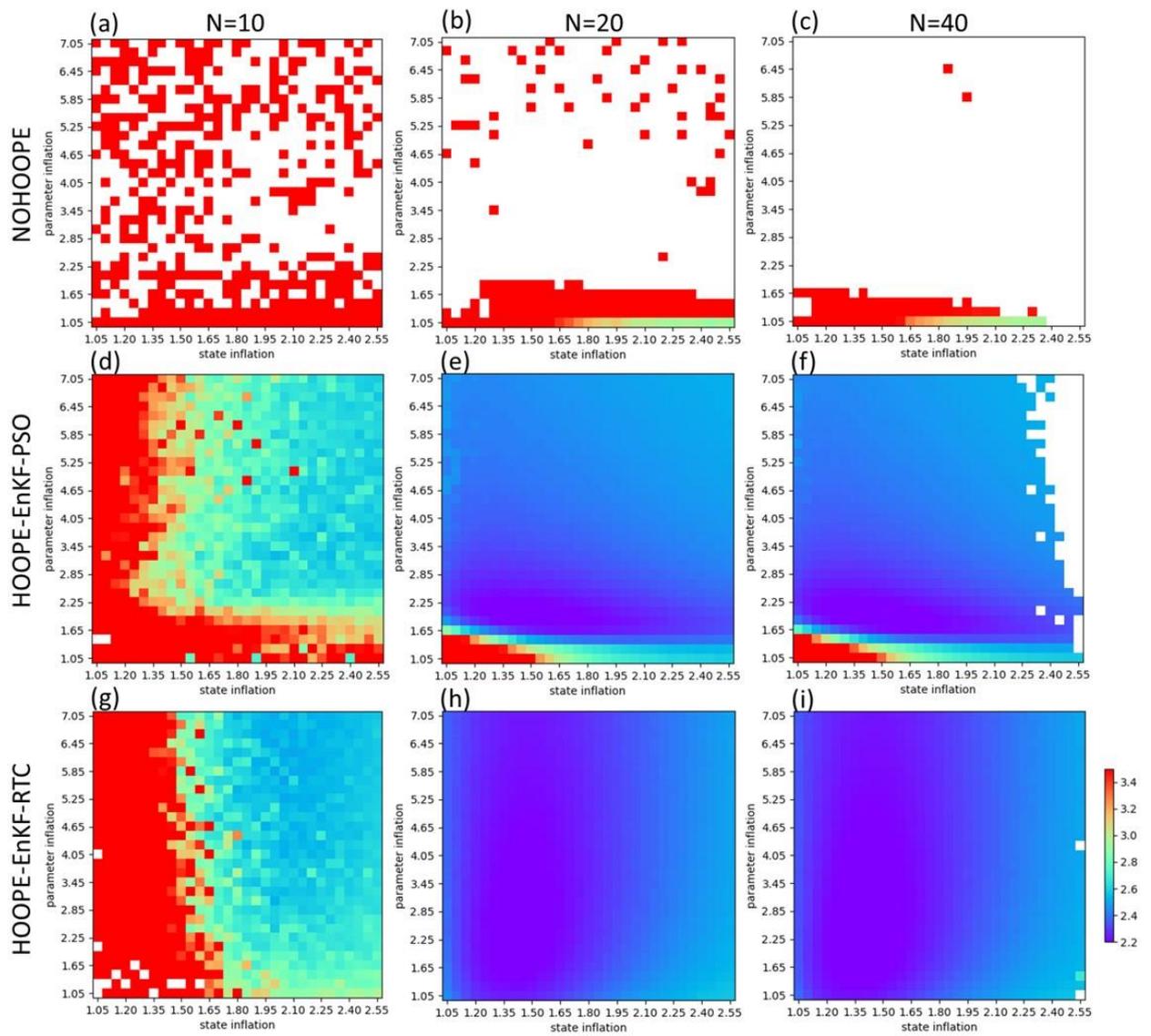

**Figure 5**. Same as Figure 3 but for RMSE of model parameters.

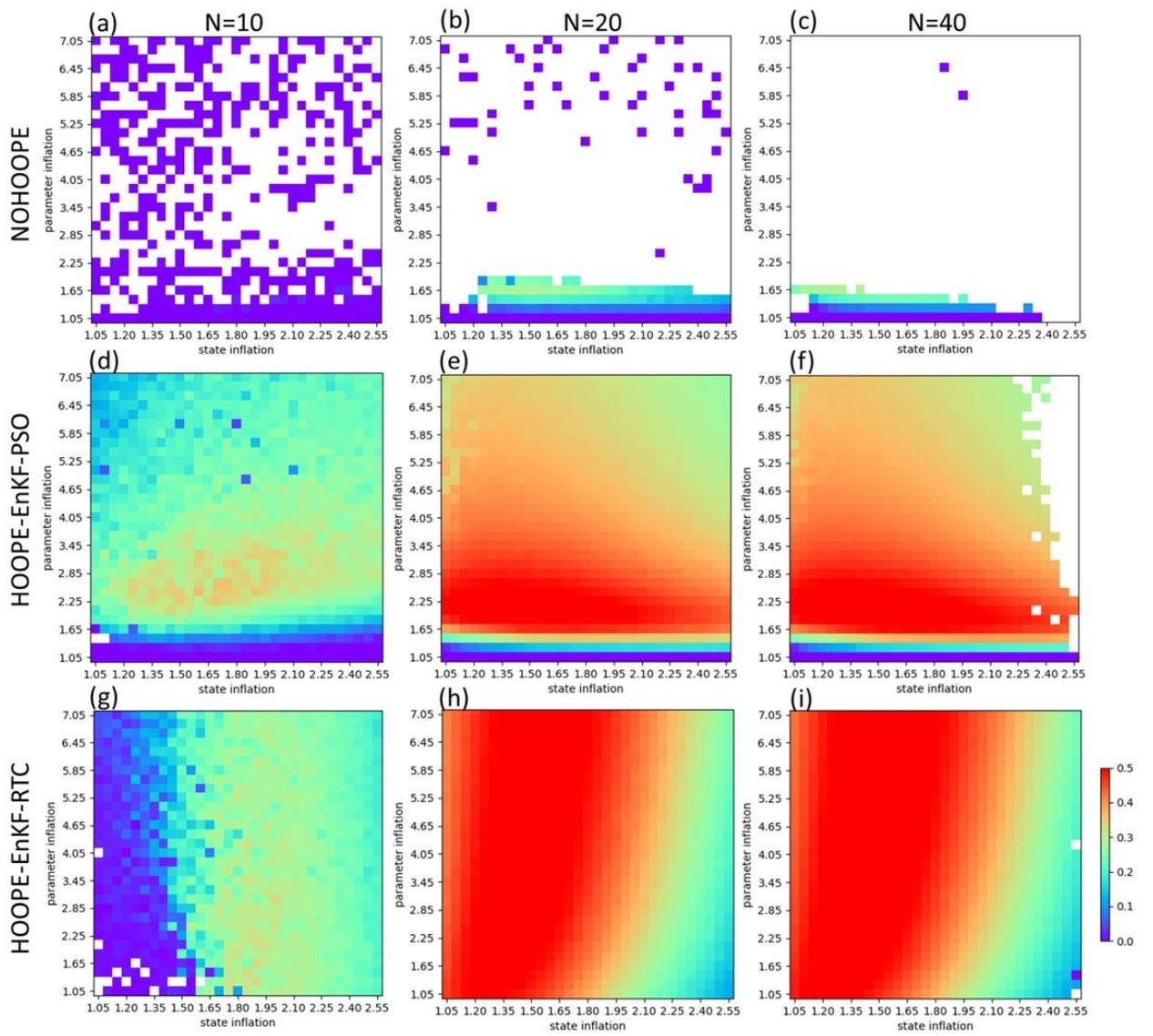

**Figure 6.** Same as Figure 3 but for R of model parameters

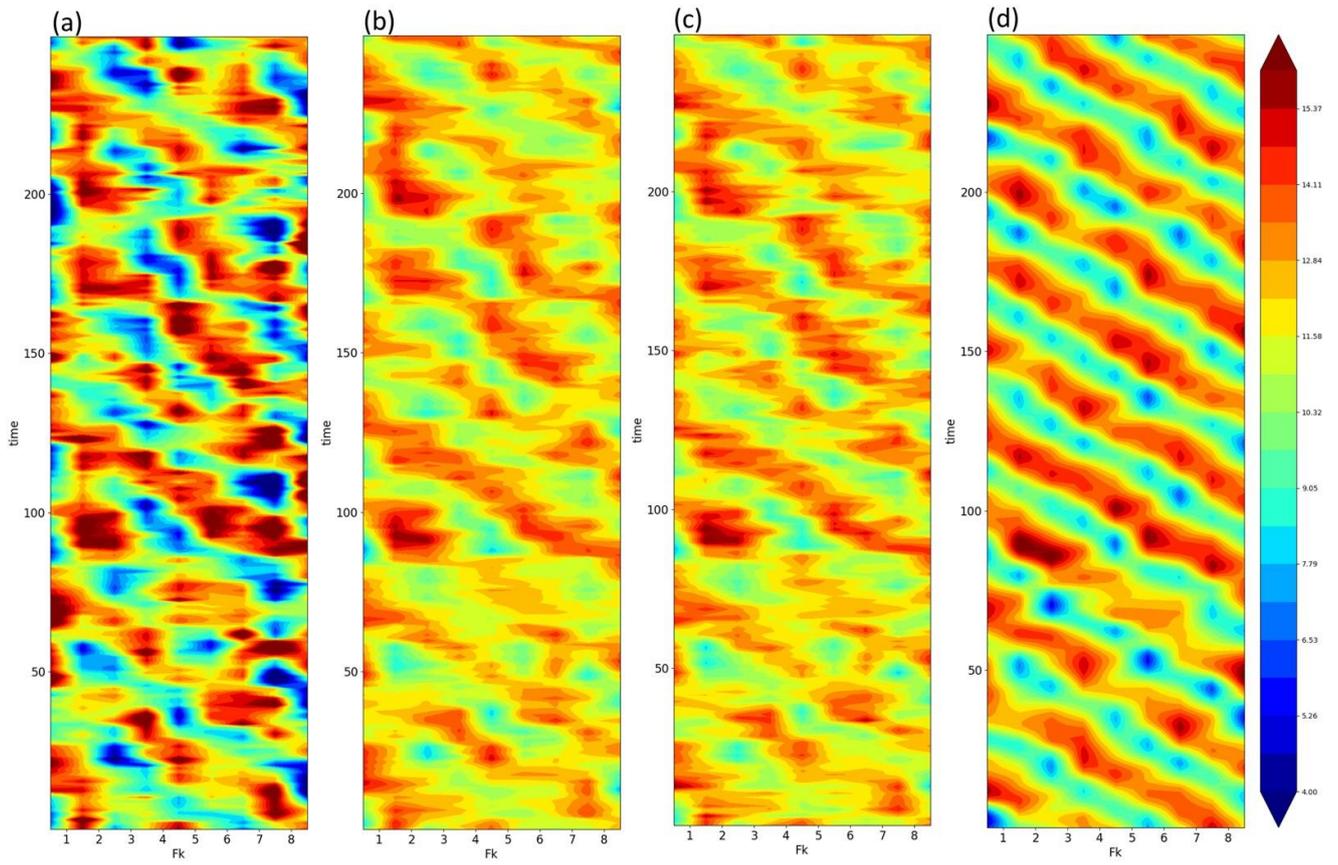

**Figure 7**. Hovmöller diagram of $F[k]$ in Equation 28 (see Section 3) estimated by mean of 40 ensemble members estimated by (a) NOHOOPE, (b) HOOPE-EnKF-PSO, (c) HOOPE-EnKF-RTC, and (d) synthetic truth. The inflation factors are adaptively determined.

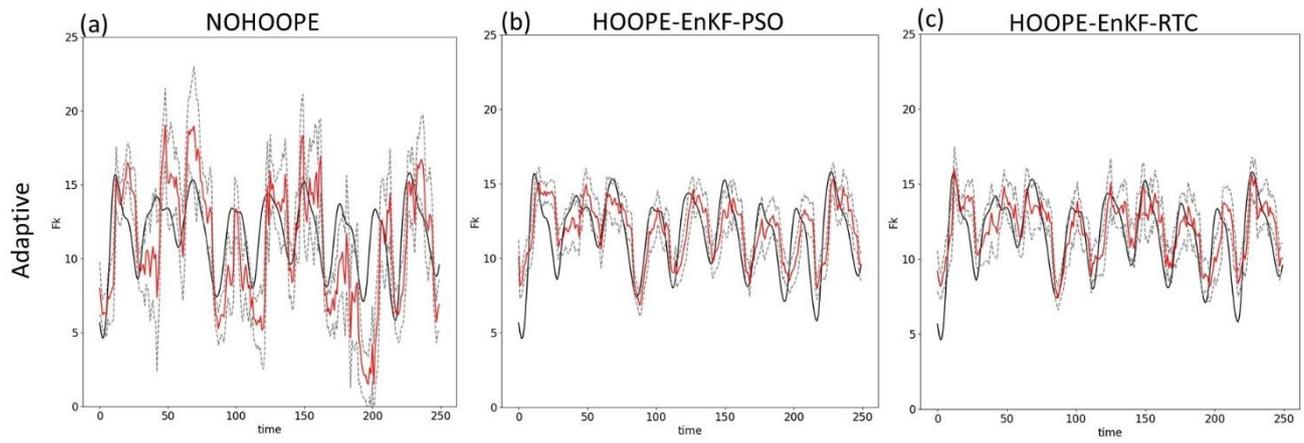

**Figure 8**. Timeseries of $F[1]$ in Equation 28 (see Section 3). Black line shows the synthetic truth. Red lines and grey dashed lines are mean and standard deviation of 40 ensemble members estimated by (a) NOHOOPE, (b) HOOPE-EnKF-PSO, and (c) HOOPE-EnKF-RTC. The inflation factors are adaptively determined.

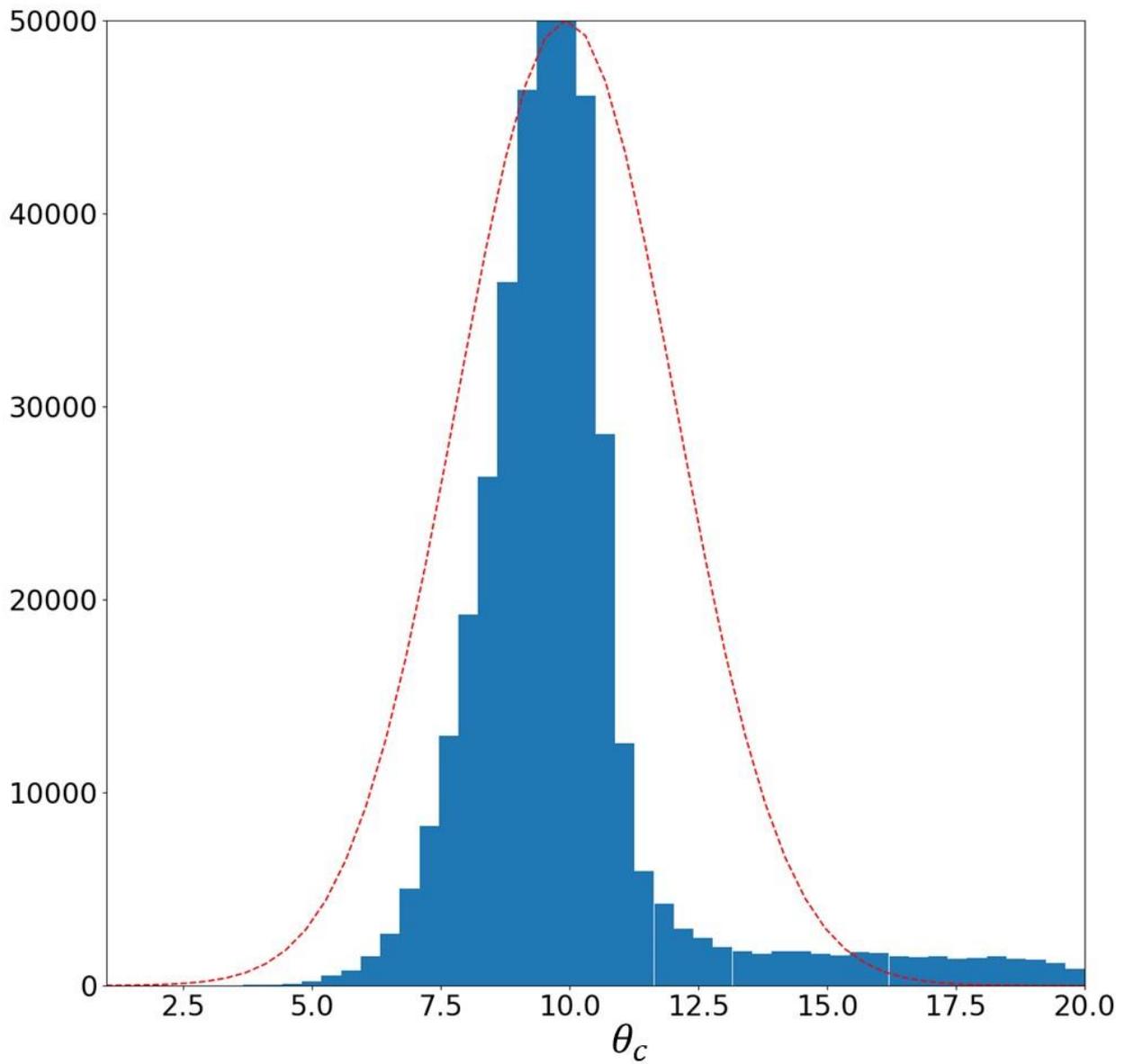

**Figure 9.** Posterior distribution of a time-invariant model parameter from (blue) samples by the Metropolis-Hastings algorithm and (red) a fitted Gaussian distribution in the SPEEDY experiment.

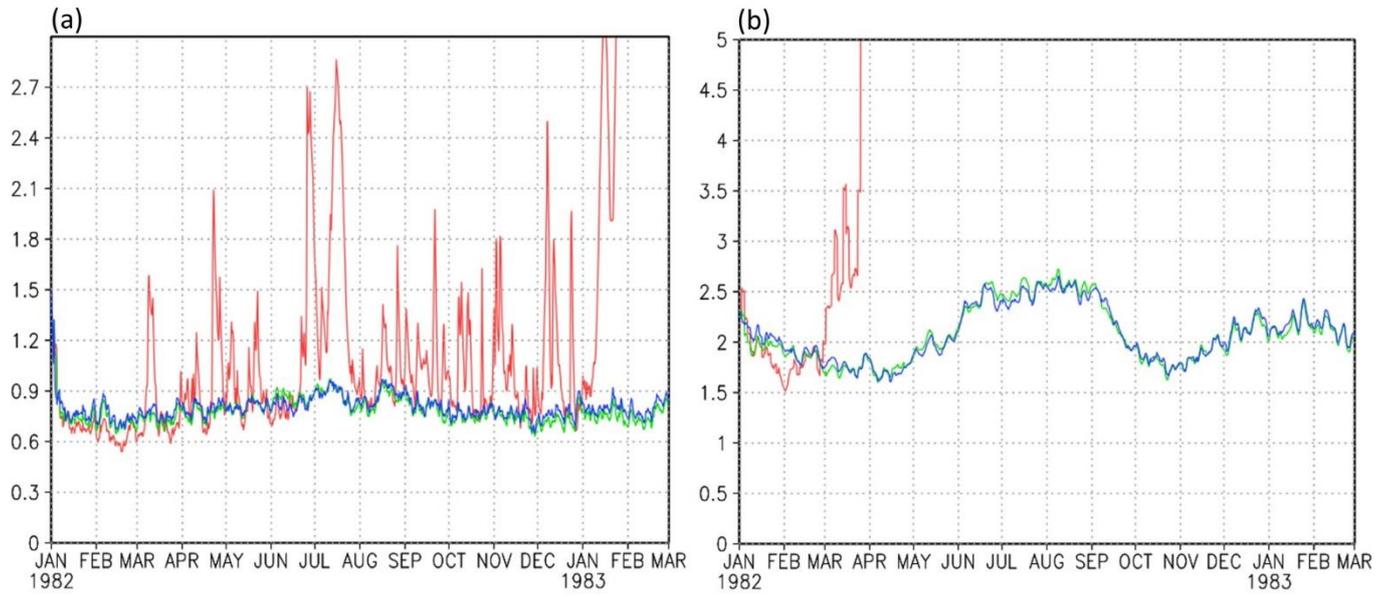

**Figure 10.** Timeseries of RMSE against the nature run of (a) specific humidity [gkg$^{-1}$] and (b) humidity threshold [gkg$^{-1}$] for convection activities in the lowest layer. Red, green, and blue lines show NOHOOPE, HOOPE-EnKF-PSO, and HOOPE-EnKF-RTC, respectively. While RMSE is calculated globally in (a), RMSE is obtained only in the tropics (30S-30N) in (b).

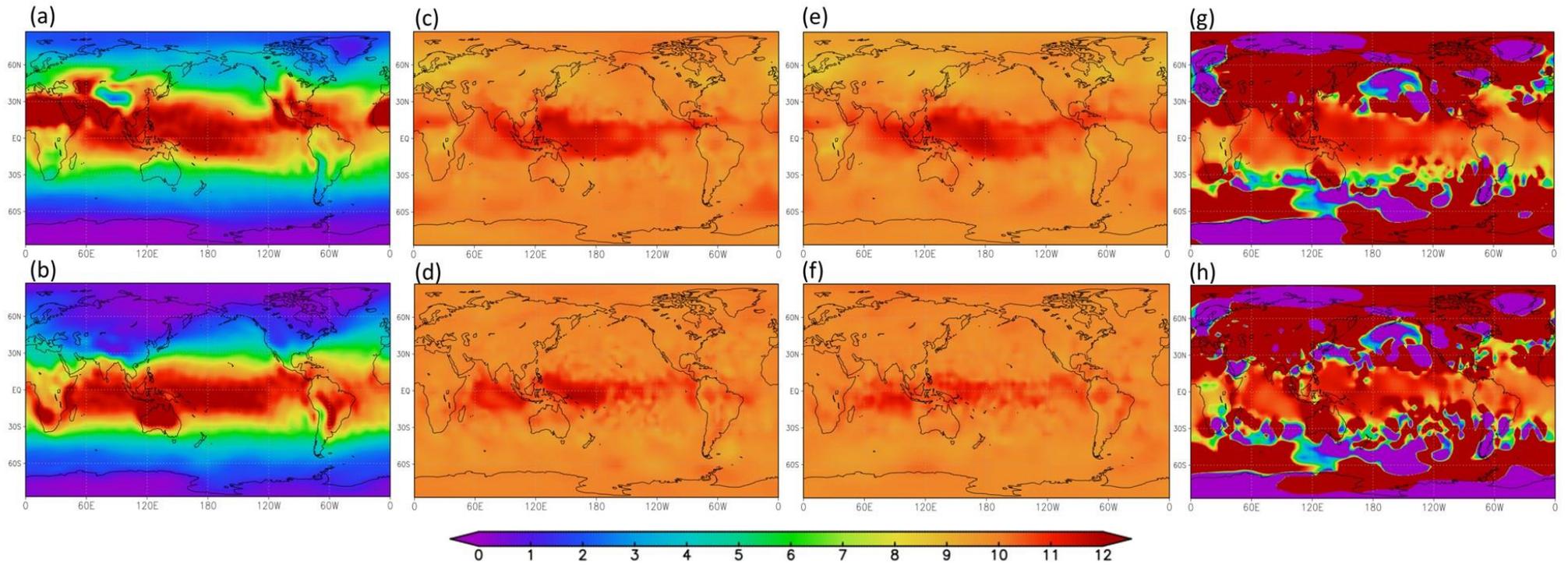

**Figure 11.** Global distribution of humidity thresholds [gkg$^{-1}$] for convection activities estimated by (a-b) the nature run, (c-d) HOOPE-EnKF-PSO, (e-f) HOOPE-EnKF-RTC, (g-h) NOHOOPE in (a,c,e,g) June-July-August 1982 and (b,d,f,h) December-January-February 1983.

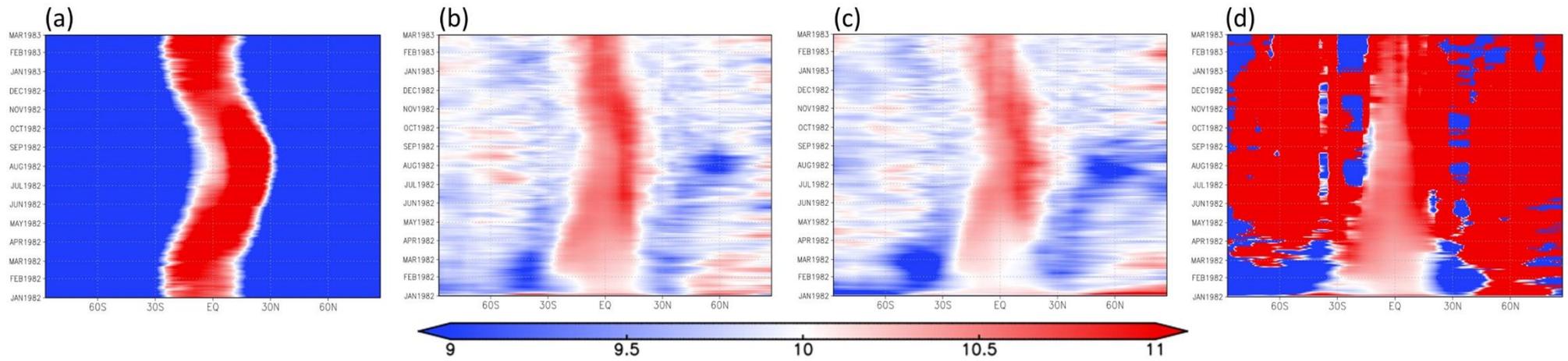

**Figure 12**. Hovmöller diagram of humidity thresholds [gkg$^{-1}$] for convection activities estimated by (a) the nature run, (b) HOOPE-EnKF-PSO, (c) HOOPE-EnKF-RTC, and (d) NOHOOPE experiments.